\documentclass[amsmath,amssymb, aps, prb, twocolumn, notitlepage]{revtex4-1}

\usepackage{graphicx}
\usepackage{dcolumn}
\usepackage{bm}
\usepackage[usenames]{color}
\usepackage{slashed}

\newcommand{\lagr}{\mathcal{L}}
\DeclareMathOperator{\Tr}{Tr}

\begin{document}

\title{Non-Abelian Fermionization and Fractional Quantum Hall Transitions
}

\author{Aaron Hui}
\affiliation{School of Applied \& Engineering Physics, Cornell University, Ithaca, New York 14853, USA}

\author{Michael Mulligan}
\affiliation{Department of Physics and Astronomy, University of California,
Riverside, CA 92511, USA}

\author{Eun-Ah Kim}
\affiliation{Department of Physics, Cornell University, Ithaca, New York 14853, USA}

\date{\today}%

\begin{abstract}
There has been a recent surge of interest in dualities relating theories of Chern-Simons gauge fields coupled to either bosons or fermions within the condensed matter community, particularly in the context of topological insulators and the half-filled Landau level. 
Here, we study the application of one such duality to the long-standing problem of quantum Hall inter-plateaux transitions. 
The key motivating experimental observations are the anomalously large value of the correlation length exponent $\nu \approx 2.3$ and that $\nu$ is observed to be super-universal, i.e., the same in the vicinity of distinct critical points [S.L. Sondhi et al., Rev. Mod. Phys. \textbf{69}, 315 (1997)].
Duality motivates effective descriptions for a fractional quantum Hall plateau transition involving a Chern-Simons field with $U(N_c)$ gauge group coupled to $N_f = 1$ fermion.
We study one class of theories in a controlled limit where $N_f \gg N_c$ and calculate $\nu$ to leading non-trivial order in the absence of disorder. 
Although these theories do not yield an anomalously large exponent $\nu$ within the large $N_f \gg N_c$ expansion, they do offer a new parameter space of theories that is apparently different from prior works involving abelian Chern-Simons gauge fields [X.-G. Wen and Y.-S. Wu, Phys. Rev. Lett. \textbf{70}, 1501 (1993); W. Chen. et al., Phys. Rev. B. \textbf{48}, 13749 (1993)].
\end{abstract}

\maketitle
\section{Introduction}
Phase transitions between different quantum Hall states have long been viewed as poster-child examples of quantum critical phenomena.\cite{Sondhi1997} 
The longitudinal resistivity $\rho_{xx}$, the width $\Delta B$ of the transition region, and $(d\rho_{xy}/dB)_\text{max}$ exhibit scaling collapse in the vicinity of the transition over almost two decades of temperature,\cite{PhysRevLett.102.216801,PhysRevB.81.033305,PhysRevLett.61.1294,PhysRevB.51.18033,PhysRevLett.74.4511,ENGEL199013} frequency,\cite{PhysRevLett.71.2638} and current.\cite{PhysRevB.50.14609}
Furthermore, although each plateau is believed to represent a distinct topologically ordered phase with (generally) different sets of fractionalized excitations, inter-plateaux transitions appear to possess the same values for the correlation length exponent $\nu\approx 2.3$ and dynamical critical exponent $z \approx 1$: distinct critical points exhibit ``super-universality.''\cite{Kivelson1992, Shimshoni1997, Sondhi1997,PhysRevLett.64.1297} 
The anomalously large value of $\nu\approx 2.3$ and the apparent super-universality remain a major mystery from the theoretical standpoint, as an accurate description clearly involves strong interactions as well as some form of translational symmetry breaking, such as disorder. 
This problem has been studied from a field-theoretic perspective using a theory of flux-attached bosons.\cite{Kivelson1992}
However it has been difficult to make progress due to the fact that the the quantum field theory of interest (matter coupled to an abelian Chern-Simons gauge field) is strongly coupled.\cite{Kivelson1992,Lopez1991,Frohlich1991,Zhang1992, Zhang1989, Sachdev1997, PhysRevB.47.7312} Controlled approximations to this theory yield correlation length exponents that strongly depend on the particular quantum Hall transition.\cite{Wen1993, Chen1993}

Duality provides a powerful perspective for studying strongly coupled quantum field theories that has been used in the past with great success.\cite{PhysRevD.11.2088, PhysRevD.11.3026, PhysRevB.9.2911, RevModPhys.52.453, 1994NuPhB.431..484S, 1996IntriligatorSeiberg, MAGOO}
There are two senses in which different theories are said to be dual.
The first is as an exact equivalence of theories.
A familiar example is bosonization in $1+1$ dimensions where a self-interacting Dirac fermion can be equivalently described by the theory of a free boson. \cite{PhysRevD.11.2088, PhysRevD.11.3026, PhysRevB.9.2911}
The second type of duality is as an IR equivalence: two theories are IR dual if they belong to the same universality class. 
In this paper, we use duality in this second sense. 
A famous example is particle-vortex duality in $2+1$ dimensions.\cite{Dasgupta1981,Peskin1978,Fisher1989a} 
This duality identifies the IR content of the XY model to that of a lattice superconductor coupled to a $U(1)$ gauge field, i.e., the Abelian-Higgs model. 
Historically, particle-vortex duality was used as a means to understand the Abelian-Higgs model, as applied to superconductivity; the XY model was relatively well understood, so duality allowed one to predict the existence of a continuous phase transition as well as its critical behavior.
Similarly, level-rank dualities were discovered, and in fact proven, for pure Chern-Simons theories.\cite{Naculich:1990hg, Nakanishi:1990hj}
As its name implies, these dualities swap the Chern-Simons level and the rank of the gauge group (in Yang-Mills regularization) up to $U(1)$ factors.\cite{Aharony2016}

Recently, generalizations of level-rank duality have been proposed.\cite{Giombi2011, 2012JHEP...03..037A, Aharony2012a, Aharony2016, Hsin:2016blu}
The conjectured duals relate theories of Chern-Simons gauge fields coupled to either fermionic or bosonic matter fields and may, in some cases, be thought of as bosonization in $2+1$ dimensions.
These dualities have been of particular interest to the condensed matter community\cite{Seiberg2016a, Karch2016, Wang2016,MrossAliceaMotrunichexplicitderivation2016} in explaining\cite{PhysRevX.5.041031, PhysRevB.93.245151,XuYou2015selfdual} the T-Pfaffian surface state of a topological insulator as well as providing a new effective description\cite{PhysRevX.5.031027} for the half-filled Landau level that is manifestly particle-hole symmetric,\cite{ PhysRevB.94.075101,PhysRevB.95.085135, PhysRevB.92.235105,Geraedtsetal2015,PhysRevLett.117.136802} thereby ``symmetrizing" the seminal work by Halperin, Lee, and Read.\cite{PhysRevB.47.7312}

We suggest that these new dualities could also be useful in understanding phase transitions between fractional quantum Hall states, as they involve theories that generalize prior effective descriptions consisting of abelian Chern-Simons gauge fields coupled to matter.\cite{Kivelson1992,Wen1993, Chen1993}

To this end, we expand in this paper upon previous efforts to understand fractional quantum Hall transitions in field theoretic models without disorder. 
In contrast to prior works,\cite{Wen1993, Chen1993} the class of theories we study consists of a Chern-Simons gauge field with non-abelian $U(N_c)$ gauge group for $N_c > 1$ coupled to $N_f$ Dirac fermions.
When $N_f = 1$, this model is dual to the theory of a fractional quantum Hall transition first studied by Wen and Wu\cite{Wen1993} and may be viewed as a generalization of the theory studied by Chen, Fisher, and Wu. \cite{Chen1993}
Although our model is strongly coupled, it can be reliably studied in various controlled limits.
In this paper, we consider the limit where $N_f \gg N_c \gg 1$.
In this large $N_f \gg N_c$ limit, we compute the correlation length exponent $\nu$ to leading non-trivial order.
Although we do not find an anomalously large $\nu$ within this expansion, effective theories with non-abelian gauge symmetry provide a larger parameter space for exploration that could yield new insights.

The remainder of this paper is organized as follows. 
In section \ref{dualities}, we write down our starting theory and discuss its fermonic dual. 
In section \ref{calc}, we discuss the calculation of the correlation length exponent $\nu$ in the fermionic theory in the large $N_f \gg N_c$ expansion. 
In section \ref{results}, we discuss our results.
An appendix contains details on the calculation of $\nu$.

\section{Dualities}
\label{dualities}

Our starting point is the field theory studied by Wen and Wu\cite{Wen1993} that describes a fractional quantum Hall to insulator transition on a lattice (without disorder) as a superfluid-Mott transition of composite bosons, tuned by the (repulsive) onsite lattice potential;\cite{Fisher1989} the phases are identified via their Hall conductivities.
When these bosons are at unit filling (appropriate to a fractional quantum Hall transition of electrons), the latter transition has an emergent relativistic symmetry.
As shown in \onlinecite{Wen1993}, such a model can be generalized to arbitrary fractional quantum Hall to fractional quantum Hall transitions by adding additional abelian gauge fields; in this paper, we choose to focus on the simplest case.
The $2+1$-dimensional Lagrangian in Euclidean signature is
\begin{align}
\label{eq:start}
    \lagr & = |(\partial_\mu - ie^* A_\mu - i a_\mu)\phi|^2 + m^2 |\phi|^2 \cr
    & + g|\phi|^4 - \frac{i}{4\pi k^B}\epsilon^{\mu\nu\lambda}a_\mu \partial_\nu a_\lambda.
\end{align}
In this theory, the fluctuating $U(1)$ Chern-Simons gauge field $a_\mu$ with $\mu \in \{1,2,3\}$ attaches $k^B$ flux quanta to the complex bosonic field $\phi$. 
These flux-attached bosons are probed by the external electromagnetic gauge field $A_\mu$ and carry charge $e^*$. 
The coupling $g$ is understood to take its IR fixed point value.
In Eq.~\eqref{eq:start}, the transition is tuned by the renormalized mass $m^2$: in the $m^2 > 0$ phase (where $\phi$ is gapped), the Hall conductivity $\sigma_{xy} = 0$; in the $m^2<0$ phase (where $\phi$ condenses), $\sigma_{xy} = -\frac{1}{k^B} \frac{(e^*)^2}{h}$; in both phases, $\sigma_{xx} = 0$ ($\sigma_{ij}$ refers to the zero-temperature dc conductivity).
For the fractional quantum Hall - Mott insulator transition, we must choose $k^B \in \mathbb{Z}$. 
For instance, to describe the $0\rightarrow 1/3$, transition, one sets $e^*=1$ and $k^B = 3$. 
We are interested in the critical properties of Eq.~\eqref{eq:start}, so we set $m^2 = 0$ for the remainder of this paper. 

We would like to study a dual description of this fractional quantum Hall to Mott insulator transition using a Chern-Simons theory with $U(N_c)$ gauge symmetry coupled to a fermion. 
For this, we need to remedy the fact that the Chern-Simons level (equal to $-1/k^B$) for $a_\mu$ in Eq.~\eqref{eq:start} is not quantized when $k^B \in \mathbb{Z}$ is greater than one 
(see footnote\footnote{One can legalize the theory by introducing a new dynamical gauge field $b$ as a constraint, giving us the Lagrangian
\begin{multline*}
   \lagr = |(\partial_\mu -ie^*A_\mu- ia_\mu)\phi|^2 + g|\phi|^4 \\ + \frac{i}{4\pi}\epsilon^{\mu\nu\lambda} \left(-k^B b_\mu\partial_\nu b_\lambda + 2 a_\mu\partial_\nu b_\lambda\right)  
\end{multline*}}). Further using a generalized particle-vortex duality,\cite{Tong2016} we arrive at 
\begin{align}
    \lagr & = |(\partial_\mu - i\hat{a}_\mu)\hat{\phi}|^2 + g|\hat{\phi}|^4 \cr
     & + \frac{i}{4\pi}\epsilon^{\mu\nu\lambda}\left(k^B \hat{a}_\mu \partial_\nu \hat{a}_\lambda + A_\mu \partial_\nu \hat{a}_\lambda\right).
    \label{eq:pv-duality}
\end{align}
Note that $\hat{\phi}$ and the $U(1)$ gauge field $\hat{a}_\mu$ in Eq.~\eqref{eq:pv-duality} are different from the corresponding fields in Eq.~\ref{eq:start}. 
A non-relativistic version of the duality between Eqs.~\eqref{eq:start} and \eqref{eq:pv-duality} was also proposed by Lee.\cite{Lee1991} 
From this point forth, we will drop the non-dynamical background gauge field $A_\mu$.

\begin{figure}[t]
\includegraphics[width = \columnwidth]{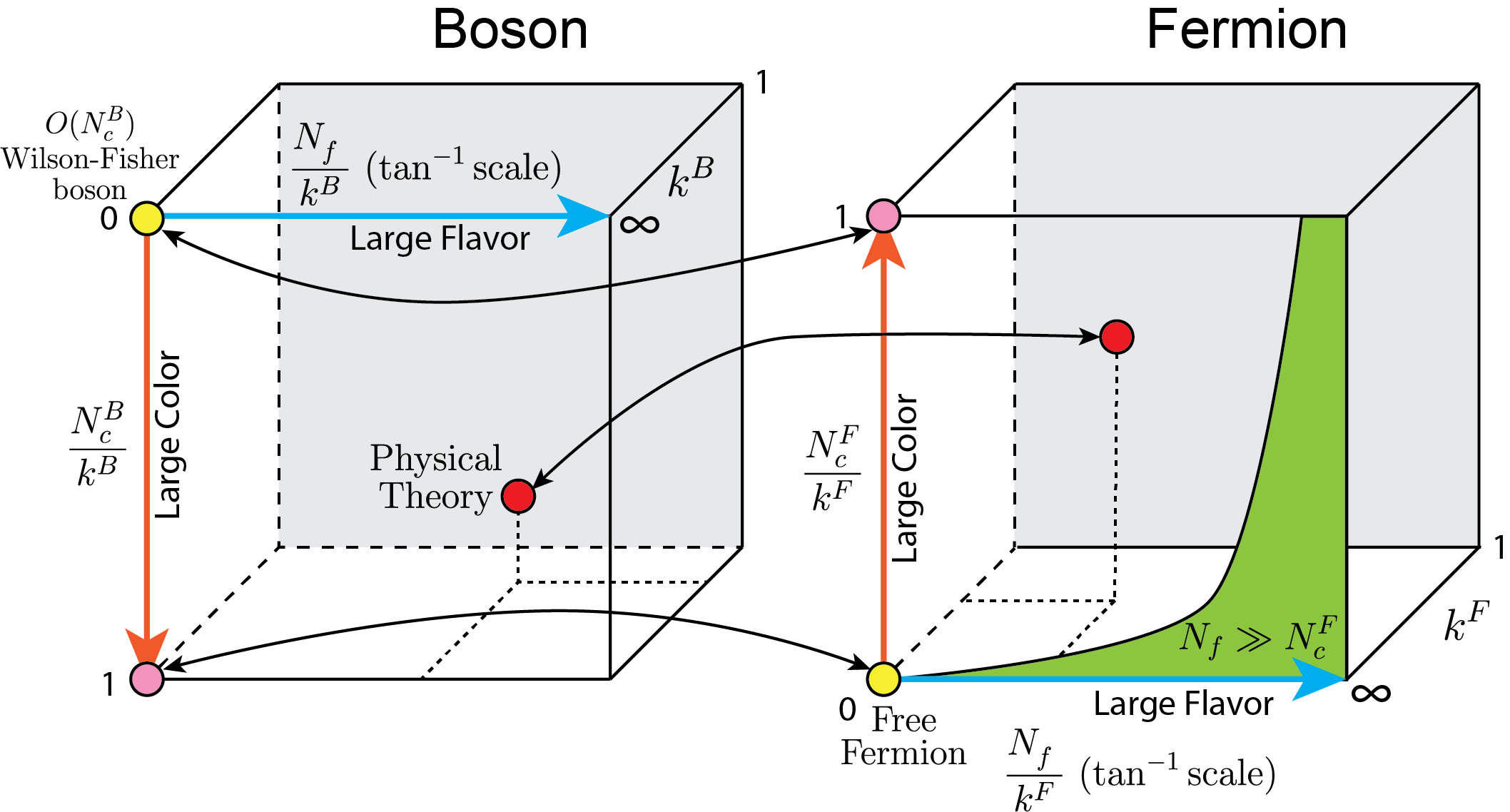}
\caption{
A schematic plot of parameter space for Chern-Simons theories with bosonic and fermionic matter. Note that the orientation of the $y$-axis is inverted between the bosonic and fermionic cubes. The double arrows indicate a duality between the connected points. The pink points refer to free theories and the yellow points to ``infinitely coupled'' theories. Previous works have studied the large color and large flavor theories both in the fermionic and bosonic cases, labeled in orange and blue.\cite{Wen1993, Chen1993, Giombi2011,Giombi2016, Aharony2012} The red dot corresponds to our physical theory, while our calculation in the $N_f \gg N_c$ expansion is done in the green region. All calculations give $\nu = 1$ at leading order,\cite{Wen1993, Chen1993, Giombi2011,Giombi2016, Aharony2012} while experiments give $\nu\approx 2.3$.\cite{Sondhi1997}}
\label{fig:parameterplot}
\end{figure}

In the hopes of understanding the effects of the strong interactions in Eq.~\eqref{eq:pv-duality}, 
we can generalize the theory in several ways: we enlarge the gauge symmetry from $U(1) \rightarrow U(N_c^B)$, where the integer $N_c^B > 1$ is the rank of the gauge group, and introduce $N_f$ flavors of bosons transforming in the fundamental representation of $U(N_c^B)$, i.e., each of the $N_f$ bosons is a vector with $N_c^B$ components.
The corresponding three-dimensional parameter space of theories is shown in the left cube in Fig.~\ref{fig:parameterplot}.
The labels for the axes are chosen to hold $N_c^B/k^B$ finite in the large $N_c^B$ limit (within the dimensional regularization scheme discussed later).
The horizontal axis is on a tan$^{-1}$ scale to make it finite in length, while the vertical axis corresponds to the 't Hooft coupling $N_c^B/k^B$, whose norm is bounded by $1$. 
The physical theory of interest with $N_f = N_c^B =1$ and $k^B\in \mathbb{Z}$ is denoted by a red dot and is located behind the front face where $k^B \rightarrow \infty$. 
Since a generic theory in Fig.~\ref{fig:parameterplot} is strongly interacting, reliable predictions are limited to small regions of the parameter space.
The best understood part is the yellow point in the top-left corner, which corresponds to the Wilson-Fisher $O(N_c^B)$ vector model, since $k^B\rightarrow \infty$ faster than $N_c^B$ and, consequently, completely suppresses the gauge fluctuations. 
In addition, large $N_f$ expansions\cite{Wen1993} (blue axis) and large $N_c^B$ expansions\cite{Giombi2016} (orange axis) have been carried out to the subleading order and leading order. 
The pink point in the bottom-left corner corresponds to ``infinite coupling,'' $N_c^B/k^B = 1$ and $k^B, N_c^B \rightarrow \infty$.

Remarkably, the recent Chern-Simons plus matter dualities sometimes relate a strongly correlated theory to a free one, and thereby constitute a non-perturbative solution to an interacting problem.
Unfortunately, this does not appear to occur for the theory described by Eq.~\eqref{eq:pv-duality}.
Instead, duality relates the IR limit of Eq.~\eqref{eq:pv-duality} to the IR limit of the theory of a Chern-Simons gauge field coupled to a Dirac fermion:
\begin{gather} 
    \lagr = |(\partial_\mu - i\hat{a}_\mu)\hat{\phi}|^2 + g|\hat{\phi}|^4 + \frac{ik^B}{4\pi}\epsilon^{\mu\nu\lambda} \hat{a}_\mu \partial_\nu \hat{a}_\lambda \nonumber \\ \updownarrow \label{eq:expanded-duality}\\ \lagr = \bar{\psi}\gamma^\mu(\partial_\mu - i\tilde{a}_\mu)\psi + \frac{ik^F}{4\pi}\epsilon^{\mu\nu\lambda}\text{Tr}\left(\tilde{a}_\mu\partial_\nu \tilde{a}_\lambda + \frac{2}{3}\tilde{a}_\mu \tilde{a}_\nu \tilde{a}_\lambda\right).\nonumber
\end{gather}
In the bottom half of \eqref{eq:expanded-duality}, $\psi$ is a 2-component fermionic field transforming in the fundamental representation of $U(k^B - 1)$, $\tilde{a}_\mu$ is a $U(k^B-1)$ gauge field, $k^F = -k^B + 1/2$, and the $\gamma$-matrices satisfy $\{\gamma^\mu, \gamma^\nu\} = 2 \delta^{\mu \nu}$. 
The trace in the non-abelian Chern-Simons term is taken with respect to the fundamental representation.
Note that we are working within dimensional regularization.\footnote{By dimensional regularization, we mean that one contracts tensor indices in 3 dimensions, while analytically continuing integrals to $3-\epsilon$ dimensions. This is sometimes called dimensional reduction in the literature.\cite{Chen1992} An alternative scheme where one regularizes with a small Yang-Mills term is equivalent to dimensional regularization, up to a constant shift of the $SU(N)$ level, so we will work exclusively in dimensional regularization for simplicity.}
See the appendix for further details.

Applying dualities\cite{Aharony2016, Hsin:2016blu} to the generalized bosonic theories with non-abelian gauge group $U(N_c^B)$ and multiple flavors $N_f$, we may schematically write:
\begin{gather}
    U(N_c^B)_{k^B, k^B} \text{ with }N_f\text{ bosons}\nonumber \\ \updownarrow \label{eq:cs-duality} \\U(k^B-N_c^B)_{-k^B+N_f/2, -k^B+N_f/2} \text{ with } N_f\text{ fermions}. \nonumber
\end{gather}
The duality in \eqref{eq:expanded-duality} is recovered by setting $N_c^B=N_f=1$. 
For the dualities in \eqref{eq:cs-duality}, the subscripts on $U(N)$ signify the levels of the $SU(N) \subset U(N)$ and $U(1) \subset U(N)$ Chern-Simons gauge fields; we will denote the rank of the gauge group in the fermionic theory of Eq.~\eqref{eq:cs-duality} with the integer $N_c^F = k^B - N_c^B$.
Armed with the dualities between generalized theories, we can now consider the three-dimensional parameter space associated with the fermionic theories (see Fig.~\ref{fig:parameterplot}).
Duality presents the choice of which representation of the same physics to study.

Fig.~\ref{fig:parameterplot} depicts the duality mappings in \eqref{eq:cs-duality}.
We denote dualities between specific points in Fig.~\ref{fig:parameterplot} with double-headed arrows that relate bosonic theories to fermionic theories. 
We intentionally chose the vertical axis of the two cubes to point in opposite directions in order to visually indicate how a strongly coupled theory on one side can map to a weakly coupled theory. 
For example, the yellow point in the bottom left corner represents the theory of a free fermion maps to an ``infinitely coupled'' bosonic theory. 
Similarly, the pink point on the top-right corner representing the ``infinitely coupled'' fermionic theory maps to the $O(N_c^B\rightarrow \infty)$ Wilson-Fisher boson. 
Unfortunately, the physical bosonic theory of interest (the red point), which is far from any known solvable point in the bosonic parameter space, maps to another strongly coupled theory on the fermionic side. 
Short of being able to directly access the physical theory, large $N_f$ expansions\cite{Chen1993} (blue axis) and large $N_c^F$ expansions\cite{Giombi2011,Giombi2016} (orange axis) have been studied on the purely fermionic side. 

In the remainder of this paper, we study the fermionic dual our physical bosonic theory (red point) using the dualities stated in \eqref{eq:expanded-duality}. 
We attempt to access this strongly coupled fermionic theory by employing a $N_f \gg N_c^F$ expansion, valid within the green region of Fig~\ref{fig:parameterplot}. 
The dualities in \eqref{eq:cs-duality} are only conjectured to hold when $N_f \leq N_c^B$\cite{Hsin:2016blu}: 
by employing the $N_f\gg N_c^F$ expansion, we are exploring a class of fermionic theories that is different from the previously studied class of bosonic theories.

\section{$N_f\gg N_c$ expansion}
\label{calc}
We generalize the fermionic side of Eq.~\eqref{eq:expanded-duality} to an arbitrary number of flavors $N_f$ so that the Lagrangian becomes
\begin{align}
    \lagr & = \sum_{i=j}^{N_f} \bar{\psi}_j\gamma^\mu(\partial_\mu - ia_\mu)\psi_j \cr
    &+ \frac{ik^F}{4\pi}\epsilon^{\mu\nu\lambda}\text{Tr}\left( a_\mu\partial_\nu a_\lambda + \frac{2}{3}a_\mu a_\nu a_\lambda\right).
    \label{eq:fermionlagrangian}
\end{align}
(We have dropped the tildes on $a$ in Eq.~\eqref{eq:fermionlagrangian}.)
The fermionic dual of the physical boson theory has $N_f =1$, $N_c^B = 1$, $N^F_c = k^B - N^B_c$ and $k^F = -k^B + N_f/2$. 

We calculate the correlation length exponent $\nu$ via the definition $\nu^{-1} = 3 - [\bar{\psi}\psi(x)]$, which comes from the fact that the correlation length $\xi \sim m^{-1}$ as the mass $m$ is the critical tuning parameter.\cite{Wen1993} 
To obtain $\nu$, we will compute the scaling dimension of the (momentum space) mass operator $J_0(p) = (\bar{\psi}\psi) (p)$. 
Recall that in position space, the scaling dimension $\delta$ is defined by the algebraic decay of $\langle J_0(x) J_0 (0) \rangle \sim x^{-2\delta}$.
Upon Fourier transforming, we have $\langle J_0(p) J_0(-p)\rangle \sim p^{2\delta - d}$, where $d = 3$ is the spacetime dimension. 
We control the calculation in the $N_f \gg N_c^F$ limit taking $k^F, N_c^F, N_f \rightarrow \infty$ while keeping the ratios $\lambda = N_f/k^F$ and $\alpha = N_c^F/N_f$ finite, along with $\alpha \ll 1$. 
Therefore, we calculate perturbatively in $\alpha$ to first subleading order and exactly in $\lambda$. 
Note that $\lambda$ can (effectively) take any value in $\mathbb{R}$ -- it is not the 't Hooft coupling $N^F_c/k^F$. 

This calculation was first investigated in a beautiful paper by Gurucharan and Prakash, where the primary motivation was to find tractable non-supersymmetric conformal field theories with gravitational duals.\cite{Gurucharan2014} 
Here, we use Eq.~\eqref{eq:fermionlagrangian} to model inter-plateaux transitions and, in the course of our study, we correct a minor error in Ref.~[\onlinecite{Gurucharan2014}].

The leading order piece $\delta^{(0)}$ of the scaling dimension of the mass operator $J_0$ in $d$ Euclidean dimensions is related to the leading order decay of the correlator by 
\begin{equation}
\langle J_0(p) J_0(-p) \rangle_\text{leading} \sim p^{2\delta^{(0)}-d},
\end{equation}
where $p$ is the momentum inserted at the $J_0$ vertex. Only the tree-level diagram contributes, which results in $\delta^{(0)} = 2$. To calculate the anomalous dimension $\delta^{(1)}$ of the mass operator $J_0$, we extract the logarithmic divergences of the 2-point correlator as in, e.g., Ref.~[\onlinecite{Chester2016}]:
\begin{equation}
\langle J_0 J_0\rangle = (1-2\delta^{(1)}\ln \Lambda + \ldots) \langle J_0 J_0 \rangle_\text{leading}.
\end{equation}
Keeping terms to $\mathcal{O}(\alpha)$, we arrive at the result:
\begin{align}
    [\bar{\psi}\psi] & = 2 - \alpha\frac{64\lambda^2}{64+\pi^2\lambda^2} \left(\frac{1}{3} + 2\cdot\frac{1}{2} \frac{64-\lambda^2\pi^2}{64+\lambda^2\pi^2}\right) \label{eq:error-compare}\\
    &= 2 - \alpha\frac{128\lambda^2}{3}\frac{128-\pi^2\lambda^2}{(64+\pi^2\lambda^2)^2} \label{eq:result}.
\end{align}
The factor of ``2" appearing before the second term in the parentheses above is the quantitative difference between our result and that in Ref.~[\onlinecite{Gurucharan2014}], and results from an additional Feynman diagram.
For calculational details, see the appendix. 
Therefore, we arrive at the result:
\begin{equation}
    \nu = 1 - \alpha\frac{128\lambda^2}{3}\frac{128-\pi^2\lambda^2}{(64+\pi^2\lambda^2)^2}.
    \label{eq:our_nu}
\end{equation}
\begin{figure}
\includegraphics[width=\columnwidth]{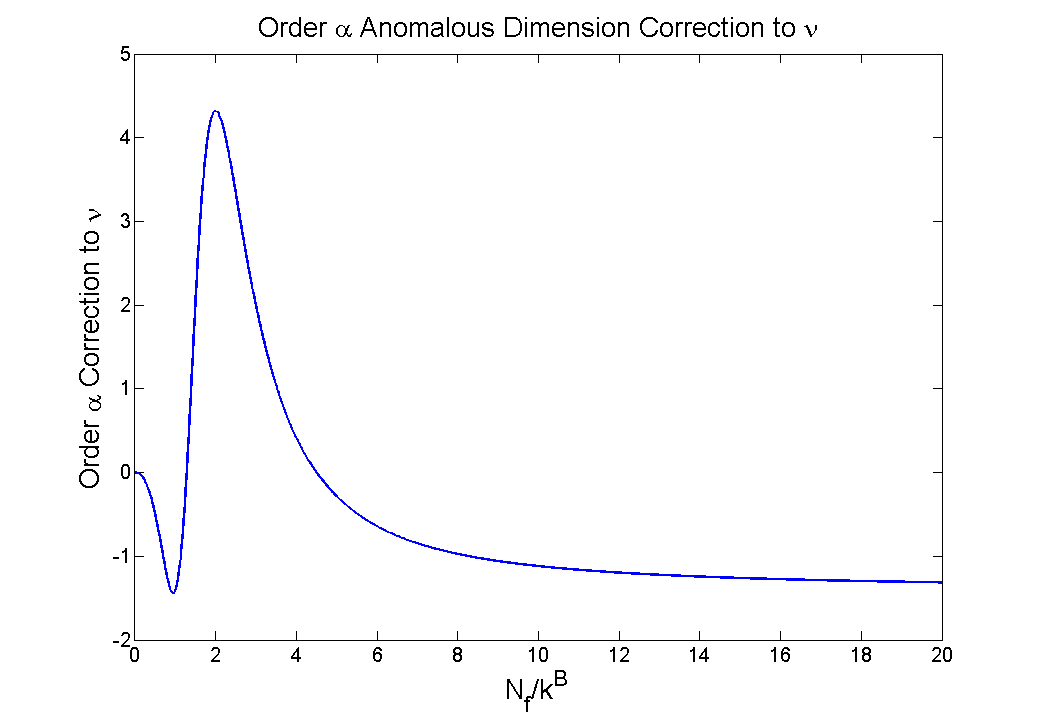}
\caption{A plot of the anomalous dimension correction to $\nu$ to $\mathcal{O}(\alpha)$ in the original bosonic parameters. The $y$ axis is in units of $\alpha$. It is positive for $1.29 < N_f/k^B < 4.50$. The parameter $\lambda$ used in Eq.~\eqref{eq:our_nu} is related to $N_f/k^B$ by $\lambda^{-1} = -k^B/N_f + 1/2$.}
\label{fig:boson-nu}
\end{figure}
We plot the anomalous dimension correction to $\nu$ at $\mathcal{O}(\alpha)$ in Fig.~\ref{fig:boson-nu} as a function of the original bosonic parameters using the relation $\lambda^{-1} = -k^B/N_f + 1/2$, with the $y$-axis measured in units of $\alpha$. 
Note that the correction is positive only when $1.29 < N_f/k^B < 4.50$. In the fermionic variables, this corresponds to $\lambda > 3.6$. 

If we want to consider the $0\rightarrow 1/3$ transition, then we should set $N_f = 1$, $N_c^B = 1$, $e^* = 1$, and $k^B = 3$.
Substituting these values into Eq.~\eqref{eq:result}, we find $\nu = 1 - .4014$. 
In this case, the correction to $\nu$ is negative. 
The dynamical critical exponent $z=1$ automatically, since our theory is Lorentz-invariant.

Chen, Fisher, and Wu studied the abelian version of Eq.~\eqref{eq:fermionlagrangian} given by
\begin{align}
    \lagr & = \sum_{i=1}^{N_f} \bar{\psi}_i\gamma^\mu(\partial_\mu - ia_\mu)\psi_i + \frac{ik^F}{4\pi}\epsilon^{\mu\nu\lambda}a_\mu\partial_\nu a_\lambda,
\end{align}
where $a_\mu$ is a $U(1)$ gauge field. 
We have rescaled $a_\mu$ to make the comparison between their theory and ours more transparent. 
They extract $\nu$ from the scaling dimension $[\bar{\psi}\psi]$ in a large $N_f$ expansion and arrive at the result,
\begin{align}
    \nu_{CFW} = 1 - \frac{1}{N_f}\frac{128\lambda^2_{CFW}}{3}\frac{128 - \pi^2\lambda^2_{CFW}}{(64+\pi^2\lambda^2_{CFW})^2},
    \label{eq:CFW_nu}
\end{align}
where $\lambda_{CFW} = N_f/k^F$. Comparing Eq.~\eqref{eq:our_nu} and Eq.~\eqref{eq:CFW_nu}, we see that the two expressions formally match.
To $\mathcal{O}(\alpha)$, our non-abelian extension to $U(N_c^F)$ only contributes an additional color factor. 
It turns out that the diagrams contributing to $\nu$ in a $N_f \gg N_c$ expansion are the same as those of a large $N_f$ expansion to subleading order, up to color factors. 
At higher orders, this equivalence is no longer expected to be true: the subleading in $N_f$ diagrams are planar because gauge lines are $1/N_f$-suppressed. 
(This formal equivalence of expansions to subleading order is likely to be true on the bosonic side as well, though we have not explicitly verified this.)
Note, however, that the two models give different results when considering a particular fractional quantum Hall transition. 
For example, in the $0\rightarrow 1/3$ transition, our model has $N_c^F = \alpha = 2$ and $k^F = -5/2$, so $\nu = 1 - .4014$. 
In the model studied by Chen, Fisher, and Wu, they set $N_f = 1$, $N_c^F = 1$, $e^* = 1/3$ and $k^F = 3/2$,\cite{Chen1993, Sachdev1997} corresponding to $``\alpha_{CFW} = 1''$ and $\lambda_{CFW} = 2/3$, so that $\nu_{CFW} = 1 - .5012$. 
Although the expressions for $\nu$ formally agree, the physical values of the parameters are different, so they should be thought of as describing different physics. \footnote{This conclusion might be further supported by the fact that the bosonization dual of the model studied by Chen, Fisher, and Wu involves a boson coupled to a Chern-Simons gauge field with non-abelian gauge group, rather than an abelian gauge field. In particular,
\begin{gather*}
    U(1)_{k^B-1/2, k^B-1/2} \text{ with one fermion} \\ 
    \protect\updownarrow \\
    U(k^B-1)_{-k^B, -k^B} \text{ with one boson}
\end{gather*}. } 

\section{Discussion}
\label{results}
The observations of super-universality and the anomalously large correlation exponent $\nu$ associated with quantum Hall inter-plateaux transitions remain a long-standing conundrum. 
Duality motivates an exploration of a large space of theories that may provide new insight.
We focused on an effective description of a fractional quantum Hall transition involving a non-abelian Chern-Simons gauge field with $U(N_c)$ gauge group and $N_f$ fermions.
This theory is dual to the critical theory of an abelian Chern-Simons gauge field coupled to a boson.
We calculated the correlation length exponent $\nu$ to first subleading order in the large $N_f \gg N_c$ expansion, filling in the green region in Fig.~\ref{fig:parameterplot}.
We found the $N_f \gg N_c^F$ expansion to be formally equivalent to a fermionic large $N_f$ expansion (blue axis) to first subleading order,\cite{Chen1993} although the precise values of the $\nu$ inferred differ.
Accordingly the exponent $\nu$ continues to depend on the pair of plateaux in question, rather than showing any super-universality. 
Moreover, the calculated exponent $\nu$ continues to be far below the experimental value.

Clearly there are many aspects of the physical problem that were left out in our model. 
It may be that translational symmetry breaking needs to be incorporated so as to include the effect of disorder. 
Also, the thus-far unexplored subleading correction in the large $N_c$ limit may prove enlightening. 
However, it appears plausible that calculating the exponent order by order with respect to some perturbative control parameter may not be the best strategy. 
Rather, it would be interesting to address the apparent super-universality in a more wholesome manner from the outset.\cite{2017arXiv171204942H} 

\section*{Acknowledgements}
We would like to thank S. Chakravarty, G.Y. Cho, M. Fisher, T. Hartman, S. Jain, S. Kivelson, M. Lawler, S. Prakash, S. Sachdev, E. Shimshoni, and S. Sondhi for helpful discussions and comments.
A.H. was supported by the National Science Foundation Graduate Research Fellowship under Grant No. DGE-1650441.
M.M. was supported in part by the UCR Academic Senate.
M.M. is grateful for the generous hospitality of the Aspen Center for Physics, which is supported by the National Science Foundation (NSF) grant PHY-1607611. 
E.-A.K. was supported by the U.S. Department of Energy, Office of Basic Energy Sciences, Division of Materials Science and Engineering under Award DE-SC0010313. E.-A.K. acknowledges Simons Fellow in Theoretical Physics Award \#392182.
The authors are also grateful for the hospitality of the Kavli Institute for Theoretical Physics, under Grant No. NSF PHY-1125915, where some of this work was performed. 
\bibliography{biblio}{}

\begin{thebibliography}{62}%
\makeatletter
\providecommand \@ifxundefined [1]{%
 \@ifx{#1\undefined}
}%
\providecommand \@ifnum [1]{%
 \ifnum #1\expandafter \@firstoftwo
 \else \expandafter \@secondoftwo
 \fi
}%
\providecommand \@ifx [1]{%
 \ifx #1\expandafter \@firstoftwo
 \else \expandafter \@secondoftwo
 \fi
}%
\providecommand \natexlab [1]{#1}%
\providecommand \enquote  [1]{``#1''}%
\providecommand \bibnamefont  [1]{#1}%
\providecommand \bibfnamefont [1]{#1}%
\providecommand \citenamefont [1]{#1}%
\providecommand \href@noop [0]{\@secondoftwo}%
\providecommand \href [0]{\begingroup \@sanitize@url \@href}%
\providecommand \@href[1]{\@@startlink{#1}\@@href}%
\providecommand \@@href[1]{\endgroup#1\@@endlink}%
\providecommand \@sanitize@url [0]{\catcode `\\12\catcode `\$12\catcode
  `\&12\catcode `\#12\catcode `\^12\catcode `\_12\catcode `\%12\relax}%
\providecommand \@@startlink[1]{}%
\providecommand \@@endlink[0]{}%
\providecommand \url  [0]{\begingroup\@sanitize@url \@url }%
\providecommand \@url [1]{\endgroup\@href {#1}{\urlprefix }}%
\providecommand \urlprefix  [0]{URL }%
\providecommand \Eprint [0]{\href }%
\providecommand \doibase [0]{http://dx.doi.org/}%
\providecommand \selectlanguage [0]{\@gobble}%
\providecommand \bibinfo  [0]{\@secondoftwo}%
\providecommand \bibfield  [0]{\@secondoftwo}%
\providecommand \translation [1]{[#1]}%
\providecommand \BibitemOpen [0]{}%
\providecommand \bibitemStop [0]{}%
\providecommand \bibitemNoStop [0]{.\EOS\space}%
\providecommand \EOS [0]{\spacefactor3000\relax}%
\providecommand \BibitemShut  [1]{\csname bibitem#1\endcsname}%
\let\auto@bib@innerbib\@empty
\bibitem [{\citenamefont {Sondhi}\ \emph {et~al.}(1997)\citenamefont {Sondhi},
  \citenamefont {Girvin}, \citenamefont {Carini},\ and\ \citenamefont
  {Shahar}}]{Sondhi1997}%
  \BibitemOpen
  \bibfield  {author} {\bibinfo {author} {\bibfnamefont {S.~L.}\ \bibnamefont
  {Sondhi}}, \bibinfo {author} {\bibfnamefont {S.~M.}\ \bibnamefont {Girvin}},
  \bibinfo {author} {\bibfnamefont {J.~P.}\ \bibnamefont {Carini}}, \ and\
  \bibinfo {author} {\bibfnamefont {D.}~\bibnamefont {Shahar}},\ }\href
  {\doibase 10.1103/RevModPhys.69.315} {\bibfield  {journal} {\bibinfo
  {journal} {Rev. Mod. Phys.}\ }\textbf {\bibinfo {volume} {69}},\ \bibinfo
  {pages} {315} (\bibinfo {year} {1997})}\BibitemShut {NoStop}%
\bibitem [{\citenamefont {Li}\ \emph {et~al.}(2009)\citenamefont {Li},
  \citenamefont {Vicente}, \citenamefont {Xia}, \citenamefont {Pan},
  \citenamefont {Tsui}, \citenamefont {Pfeiffer},\ and\ \citenamefont
  {West}}]{PhysRevLett.102.216801}%
  \BibitemOpen
  \bibfield  {author} {\bibinfo {author} {\bibfnamefont {W.}~\bibnamefont
  {Li}}, \bibinfo {author} {\bibfnamefont {C.~L.}\ \bibnamefont {Vicente}},
  \bibinfo {author} {\bibfnamefont {J.~S.}\ \bibnamefont {Xia}}, \bibinfo
  {author} {\bibfnamefont {W.}~\bibnamefont {Pan}}, \bibinfo {author}
  {\bibfnamefont {D.~C.}\ \bibnamefont {Tsui}}, \bibinfo {author}
  {\bibfnamefont {L.~N.}\ \bibnamefont {Pfeiffer}}, \ and\ \bibinfo {author}
  {\bibfnamefont {K.~W.}\ \bibnamefont {West}},\ }\href {\doibase
  10.1103/PhysRevLett.102.216801} {\bibfield  {journal} {\bibinfo  {journal}
  {Phys. Rev. Lett.}\ }\textbf {\bibinfo {volume} {102}},\ \bibinfo {pages}
  {216801} (\bibinfo {year} {2009})}\BibitemShut {NoStop}%
\bibitem [{\citenamefont {Li}\ \emph {et~al.}(2010)\citenamefont {Li},
  \citenamefont {Xia}, \citenamefont {Vicente}, \citenamefont {Sullivan},
  \citenamefont {Pan}, \citenamefont {Tsui}, \citenamefont {Pfeiffer},\ and\
  \citenamefont {West}}]{PhysRevB.81.033305}%
  \BibitemOpen
  \bibfield  {author} {\bibinfo {author} {\bibfnamefont {W.}~\bibnamefont
  {Li}}, \bibinfo {author} {\bibfnamefont {J.~S.}\ \bibnamefont {Xia}},
  \bibinfo {author} {\bibfnamefont {C.}~\bibnamefont {Vicente}}, \bibinfo
  {author} {\bibfnamefont {N.~S.}\ \bibnamefont {Sullivan}}, \bibinfo {author}
  {\bibfnamefont {W.}~\bibnamefont {Pan}}, \bibinfo {author} {\bibfnamefont
  {D.~C.}\ \bibnamefont {Tsui}}, \bibinfo {author} {\bibfnamefont {L.~N.}\
  \bibnamefont {Pfeiffer}}, \ and\ \bibinfo {author} {\bibfnamefont {K.~W.}\
  \bibnamefont {West}},\ }\href {\doibase 10.1103/PhysRevB.81.033305}
  {\bibfield  {journal} {\bibinfo  {journal} {Phys. Rev. B}\ }\textbf {\bibinfo
  {volume} {81}},\ \bibinfo {pages} {033305} (\bibinfo {year}
  {2010})}\BibitemShut {NoStop}%
\bibitem [{\citenamefont {Wei}\ \emph {et~al.}(1988)\citenamefont {Wei},
  \citenamefont {Tsui}, \citenamefont {Paalanen},\ and\ \citenamefont
  {Pruisken}}]{PhysRevLett.61.1294}%
  \BibitemOpen
  \bibfield  {author} {\bibinfo {author} {\bibfnamefont {H.~P.}\ \bibnamefont
  {Wei}}, \bibinfo {author} {\bibfnamefont {D.~C.}\ \bibnamefont {Tsui}},
  \bibinfo {author} {\bibfnamefont {M.~A.}\ \bibnamefont {Paalanen}}, \ and\
  \bibinfo {author} {\bibfnamefont {A.~M.~M.}\ \bibnamefont {Pruisken}},\
  }\href {\doibase 10.1103/PhysRevLett.61.1294} {\bibfield  {journal} {\bibinfo
   {journal} {Phys. Rev. Lett.}\ }\textbf {\bibinfo {volume} {61}},\ \bibinfo
  {pages} {1294} (\bibinfo {year} {1988})}\BibitemShut {NoStop}%
\bibitem [{\citenamefont {Wong}\ \emph {et~al.}(1995)\citenamefont {Wong},
  \citenamefont {Jiang}, \citenamefont {Trivedi},\ and\ \citenamefont
  {Palm}}]{PhysRevB.51.18033}%
  \BibitemOpen
  \bibfield  {author} {\bibinfo {author} {\bibfnamefont {L.~W.}\ \bibnamefont
  {Wong}}, \bibinfo {author} {\bibfnamefont {H.~W.}\ \bibnamefont {Jiang}},
  \bibinfo {author} {\bibfnamefont {N.}~\bibnamefont {Trivedi}}, \ and\
  \bibinfo {author} {\bibfnamefont {E.}~\bibnamefont {Palm}},\ }\href {\doibase
  10.1103/PhysRevB.51.18033} {\bibfield  {journal} {\bibinfo  {journal} {Phys.
  Rev. B}\ }\textbf {\bibinfo {volume} {51}},\ \bibinfo {pages} {18033}
  (\bibinfo {year} {1995})}\BibitemShut {NoStop}%
\bibitem [{\citenamefont {Shahar}\ \emph {et~al.}(1995)\citenamefont {Shahar},
  \citenamefont {Tsui}, \citenamefont {Shayegan}, \citenamefont {Bhatt},\ and\
  \citenamefont {Cunningham}}]{PhysRevLett.74.4511}%
  \BibitemOpen
  \bibfield  {author} {\bibinfo {author} {\bibfnamefont {D.}~\bibnamefont
  {Shahar}}, \bibinfo {author} {\bibfnamefont {D.~C.}\ \bibnamefont {Tsui}},
  \bibinfo {author} {\bibfnamefont {M.}~\bibnamefont {Shayegan}}, \bibinfo
  {author} {\bibfnamefont {R.~N.}\ \bibnamefont {Bhatt}}, \ and\ \bibinfo
  {author} {\bibfnamefont {J.~E.}\ \bibnamefont {Cunningham}},\ }\href
  {\doibase 10.1103/PhysRevLett.74.4511} {\bibfield  {journal} {\bibinfo
  {journal} {Phys. Rev. Lett.}\ }\textbf {\bibinfo {volume} {74}},\ \bibinfo
  {pages} {4511} (\bibinfo {year} {1995})}\BibitemShut {NoStop}%
\bibitem [{\citenamefont {Engel}\ \emph {et~al.}(1990)\citenamefont {Engel},
  \citenamefont {Wei}, \citenamefont {Tsui},\ and\ \citenamefont
  {Shayegan}}]{ENGEL199013}%
  \BibitemOpen
  \bibfield  {author} {\bibinfo {author} {\bibfnamefont {L.}~\bibnamefont
  {Engel}}, \bibinfo {author} {\bibfnamefont {H.}~\bibnamefont {Wei}}, \bibinfo
  {author} {\bibfnamefont {D.}~\bibnamefont {Tsui}}, \ and\ \bibinfo {author}
  {\bibfnamefont {M.}~\bibnamefont {Shayegan}},\ }\href {\doibase
  https://doi.org/10.1016/0039-6028(90)90820-X} {\bibfield  {journal} {\bibinfo
   {journal} {Surf. Sci.}\ }\textbf {\bibinfo {volume} {229}},\ \bibinfo
  {pages} {13 } (\bibinfo {year} {1990})}\BibitemShut {NoStop}%
\bibitem [{\citenamefont {Engel}\ \emph {et~al.}(1993)\citenamefont {Engel},
  \citenamefont {Shahar}, \citenamefont {Kurdak},\ and\ \citenamefont
  {Tsui}}]{PhysRevLett.71.2638}%
  \BibitemOpen
  \bibfield  {author} {\bibinfo {author} {\bibfnamefont {L.~W.}\ \bibnamefont
  {Engel}}, \bibinfo {author} {\bibfnamefont {D.}~\bibnamefont {Shahar}},
  \bibinfo {author} {\bibfnamefont {{\ifmmode {\mbox{\c{C}}}\else
  {\c{C}}\fi}.}~\bibnamefont {Kurdak}}, \ and\ \bibinfo {author} {\bibfnamefont
  {D.~C.}\ \bibnamefont {Tsui}},\ }\href {\doibase 10.1103/PhysRevLett.71.2638}
  {\bibfield  {journal} {\bibinfo  {journal} {Phys. Rev. Lett.}\ }\textbf
  {\bibinfo {volume} {71}},\ \bibinfo {pages} {2638} (\bibinfo {year}
  {1993})}\BibitemShut {NoStop}%
\bibitem [{\citenamefont {Wei}\ \emph {et~al.}(1994)\citenamefont {Wei},
  \citenamefont {Engel},\ and\ \citenamefont {Tsui}}]{PhysRevB.50.14609}%
  \BibitemOpen
  \bibfield  {author} {\bibinfo {author} {\bibfnamefont {H.~P.}\ \bibnamefont
  {Wei}}, \bibinfo {author} {\bibfnamefont {L.~W.}\ \bibnamefont {Engel}}, \
  and\ \bibinfo {author} {\bibfnamefont {D.~C.}\ \bibnamefont {Tsui}},\ }\href
  {\doibase 10.1103/PhysRevB.50.14609} {\bibfield  {journal} {\bibinfo
  {journal} {Phys. Rev. B}\ }\textbf {\bibinfo {volume} {50}},\ \bibinfo
  {pages} {14609} (\bibinfo {year} {1994})}\BibitemShut {NoStop}%
\bibitem [{\citenamefont {Kivelson}\ \emph {et~al.}(1992)\citenamefont
  {Kivelson}, \citenamefont {Lee},\ and\ \citenamefont {Zhang}}]{Kivelson1992}%
  \BibitemOpen
  \bibfield  {author} {\bibinfo {author} {\bibfnamefont {S.}~\bibnamefont
  {Kivelson}}, \bibinfo {author} {\bibfnamefont {D.-H.}\ \bibnamefont {Lee}}, \
  and\ \bibinfo {author} {\bibfnamefont {S.-C.}\ \bibnamefont {Zhang}},\ }\href
  {\doibase 10.1103/PhysRevB.46.2223} {\bibfield  {journal} {\bibinfo
  {journal} {Phys. Rev. B.}\ }\textbf {\bibinfo {volume} {46}},\ \bibinfo
  {pages} {2223} (\bibinfo {year} {1992})}\BibitemShut {NoStop}%
\bibitem [{\citenamefont {Shimshoni}\ \emph {et~al.}(1997)\citenamefont
  {Shimshoni}, \citenamefont {Sondhi},\ and\ \citenamefont
  {Shahar}}]{Shimshoni1997}%
  \BibitemOpen
  \bibfield  {author} {\bibinfo {author} {\bibfnamefont {E.}~\bibnamefont
  {Shimshoni}}, \bibinfo {author} {\bibfnamefont {S.~L.}\ \bibnamefont
  {Sondhi}}, \ and\ \bibinfo {author} {\bibfnamefont {D.}~\bibnamefont
  {Shahar}},\ }\href {\doibase 10.1103/PhysRevB.55.13730} {\bibfield  {journal}
  {\bibinfo  {journal} {Phys. Rev. B}\ }\textbf {\bibinfo {volume} {55}},\
  \bibinfo {pages} {13730} (\bibinfo {year} {1997})},\ \Eprint
  {http://arxiv.org/abs/9610102} {arXiv:9610102 [cond-mat]} \BibitemShut
  {NoStop}%
\bibitem [{\citenamefont {Jain}\ \emph {et~al.}(1990)\citenamefont {Jain},
  \citenamefont {Kivelson},\ and\ \citenamefont
  {Trivedi}}]{PhysRevLett.64.1297}%
  \BibitemOpen
  \bibfield  {author} {\bibinfo {author} {\bibfnamefont {J.~K.}\ \bibnamefont
  {Jain}}, \bibinfo {author} {\bibfnamefont {S.~A.}\ \bibnamefont {Kivelson}},
  \ and\ \bibinfo {author} {\bibfnamefont {N.}~\bibnamefont {Trivedi}},\ }\href
  {\doibase 10.1103/PhysRevLett.64.1297} {\bibfield  {journal} {\bibinfo
  {journal} {Phys. Rev. Lett.}\ }\textbf {\bibinfo {volume} {64}},\ \bibinfo
  {pages} {1297} (\bibinfo {year} {1990})}\BibitemShut {NoStop}%
\bibitem [{\citenamefont {Lopez}\ and\ \citenamefont
  {Fradkin}(1991)}]{Lopez1991}%
  \BibitemOpen
  \bibfield  {author} {\bibinfo {author} {\bibfnamefont {A.}~\bibnamefont
  {Lopez}}\ and\ \bibinfo {author} {\bibfnamefont {E.}~\bibnamefont
  {Fradkin}},\ }\href {\doibase 10.1103/PhysRevB.44.5246} {\bibfield  {journal}
  {\bibinfo  {journal} {Phys. Rev. B.}\ }\textbf {\bibinfo {volume} {44}},\
  \bibinfo {pages} {5246} (\bibinfo {year} {1991})}\BibitemShut {NoStop}%
\bibitem [{\citenamefont {Fr{\"{o}}hlich}\ and\ \citenamefont
  {Zee}(1991)}]{Frohlich1991}%
  \BibitemOpen
  \bibfield  {author} {\bibinfo {author} {\bibfnamefont {J.}~\bibnamefont
  {Fr{\"{o}}hlich}}\ and\ \bibinfo {author} {\bibfnamefont {A.}~\bibnamefont
  {Zee}},\ }\href {\doibase 10.1016/0550-3213(91)90275-3} {\bibfield  {journal}
  {\bibinfo  {journal} {Nuclear Physics B}\ }\textbf {\bibinfo {volume}
  {364}},\ \bibinfo {pages} {517} (\bibinfo {year} {1991})}\BibitemShut
  {NoStop}%
\bibitem [{\citenamefont {Zhang}(1992)}]{Zhang1992}%
  \BibitemOpen
  \bibfield  {author} {\bibinfo {author} {\bibfnamefont {S.~C.}\ \bibnamefont
  {Zhang}},\ }\href {\doibase 10.1142/S0217979292000499} {\bibfield  {journal}
  {\bibinfo  {journal} {International Journal of Modern Physics B}\ }\textbf
  {\bibinfo {volume} {06}},\ \bibinfo {pages} {803} (\bibinfo {year}
  {1992})}\BibitemShut {NoStop}%
\bibitem [{\citenamefont {Zhang}\ \emph {et~al.}(1989)\citenamefont {Zhang},
  \citenamefont {Hansson},\ and\ \citenamefont {Kivelson}}]{Zhang1989}%
  \BibitemOpen
  \bibfield  {author} {\bibinfo {author} {\bibfnamefont {S.~C.}\ \bibnamefont
  {Zhang}}, \bibinfo {author} {\bibfnamefont {T.~H.}\ \bibnamefont {Hansson}},
  \ and\ \bibinfo {author} {\bibfnamefont {S.}~\bibnamefont {Kivelson}},\
  }\href {\doibase 10.1103/PhysRevLett.62.980.3} {\bibfield  {journal}
  {\bibinfo  {journal} {Phys. Rev. Lett.}\ }\textbf {\bibinfo {volume} {62}},\
  \bibinfo {pages} {980} (\bibinfo {year} {1989})}\BibitemShut {NoStop}%
\bibitem [{\citenamefont {Sachdev}(1998)}]{Sachdev1997}%
  \BibitemOpen
  \bibfield  {author} {\bibinfo {author} {\bibfnamefont {S.}~\bibnamefont
  {Sachdev}},\ }\href {\doibase 10.1103/PhysRevB.57.7157} {\bibfield  {journal}
  {\bibinfo  {journal} {Phys. Rev. B}\ }\textbf {\bibinfo {volume} {57}},\
  \bibinfo {pages} {7157} (\bibinfo {year} {1998})},\ \Eprint
  {http://arxiv.org/abs/9709243} {arXiv:9709243 [cond-mat]} \BibitemShut
  {NoStop}%
\bibitem [{\citenamefont {Halperin}\ \emph {et~al.}(1993)\citenamefont
  {Halperin}, \citenamefont {Lee},\ and\ \citenamefont
  {Read}}]{PhysRevB.47.7312}%
  \BibitemOpen
  \bibfield  {author} {\bibinfo {author} {\bibfnamefont {B.~I.}\ \bibnamefont
  {Halperin}}, \bibinfo {author} {\bibfnamefont {P.~A.}\ \bibnamefont {Lee}}, \
  and\ \bibinfo {author} {\bibfnamefont {N.}~\bibnamefont {Read}},\ }\href
  {\doibase 10.1103/PhysRevB.47.7312} {\bibfield  {journal} {\bibinfo
  {journal} {Phys. Rev. B}\ }\textbf {\bibinfo {volume} {47}},\ \bibinfo
  {pages} {7312} (\bibinfo {year} {1993})}\BibitemShut {NoStop}%
\bibitem [{\citenamefont {Wen}\ and\ \citenamefont {Wu}(1993)}]{Wen1993}%
  \BibitemOpen
  \bibfield  {author} {\bibinfo {author} {\bibfnamefont {X.-G.}\ \bibnamefont
  {Wen}}\ and\ \bibinfo {author} {\bibfnamefont {Y.-S.}\ \bibnamefont {Wu}},\
  }\href {\doibase 10.1103/PhysRevLett.70.1501} {\bibfield  {journal} {\bibinfo
   {journal} {Phys. Rev. Lett.}\ }\textbf {\bibinfo {volume} {70}},\ \bibinfo
  {pages} {1501} (\bibinfo {year} {1993})}\BibitemShut {NoStop}%
\bibitem [{\citenamefont {Chen}\ \emph {et~al.}(1993)\citenamefont {Chen},
  \citenamefont {Fisher},\ and\ \citenamefont {Wu}}]{Chen1993}%
  \BibitemOpen
  \bibfield  {author} {\bibinfo {author} {\bibfnamefont {W.}~\bibnamefont
  {Chen}}, \bibinfo {author} {\bibfnamefont {M.~P.~A.}\ \bibnamefont {Fisher}},
  \ and\ \bibinfo {author} {\bibfnamefont {Y.-S.}\ \bibnamefont {Wu}},\ }\href
  {\doibase 10.1103/PhysRevB.48.13749} {\bibfield  {journal} {\bibinfo
  {journal} {Phys. Rev. B.}\ }\textbf {\bibinfo {volume} {48}},\ \bibinfo
  {pages} {13749} (\bibinfo {year} {1993})},\ \Eprint
  {http://arxiv.org/abs/9301037} {arXiv:9301037 [cond-mat]} \BibitemShut
  {NoStop}%
\bibitem [{\citenamefont {Coleman}(1975)}]{PhysRevD.11.2088}%
  \BibitemOpen
  \bibfield  {author} {\bibinfo {author} {\bibfnamefont {S.}~\bibnamefont
  {Coleman}},\ }\href {\doibase 10.1103/PhysRevD.11.2088} {\bibfield  {journal}
  {\bibinfo  {journal} {Phys. Rev. D}\ }\textbf {\bibinfo {volume} {11}},\
  \bibinfo {pages} {2088} (\bibinfo {year} {1975})}\BibitemShut {NoStop}%
\bibitem [{\citenamefont {Mandelstam}(1975)}]{PhysRevD.11.3026}%
  \BibitemOpen
  \bibfield  {author} {\bibinfo {author} {\bibfnamefont {S.}~\bibnamefont
  {Mandelstam}},\ }\href {\doibase 10.1103/PhysRevD.11.3026} {\bibfield
  {journal} {\bibinfo  {journal} {Phys. Rev. D}\ }\textbf {\bibinfo {volume}
  {11}},\ \bibinfo {pages} {3026} (\bibinfo {year} {1975})}\BibitemShut
  {NoStop}%
\bibitem [{\citenamefont {Luther}\ and\ \citenamefont
  {Peschel}(1974)}]{PhysRevB.9.2911}%
  \BibitemOpen
  \bibfield  {author} {\bibinfo {author} {\bibfnamefont {A.}~\bibnamefont
  {Luther}}\ and\ \bibinfo {author} {\bibfnamefont {I.}~\bibnamefont
  {Peschel}},\ }\href {\doibase 10.1103/PhysRevB.9.2911} {\bibfield  {journal}
  {\bibinfo  {journal} {Phys. Rev. B}\ }\textbf {\bibinfo {volume} {9}},\
  \bibinfo {pages} {2911} (\bibinfo {year} {1974})}\BibitemShut {NoStop}%
\bibitem [{\citenamefont {Savit}(1980)}]{RevModPhys.52.453}%
  \BibitemOpen
  \bibfield  {author} {\bibinfo {author} {\bibfnamefont {R.}~\bibnamefont
  {Savit}},\ }\href {\doibase 10.1103/RevModPhys.52.453} {\bibfield  {journal}
  {\bibinfo  {journal} {Rev. Mod. Phys.}\ }\textbf {\bibinfo {volume} {52}},\
  \bibinfo {pages} {453} (\bibinfo {year} {1980})}\BibitemShut {NoStop}%
\bibitem [{\citenamefont {{Seiberg}}\ and\ \citenamefont
  {{Witten}}(1994)}]{1994NuPhB.431..484S}%
  \BibitemOpen
  \bibfield  {author} {\bibinfo {author} {\bibfnamefont {N.}~\bibnamefont
  {{Seiberg}}}\ and\ \bibinfo {author} {\bibfnamefont {E.}~\bibnamefont
  {{Witten}}},\ }\href {\doibase 10.1016/0550-3213(94)90214-3} {\bibfield
  {journal} {\bibinfo  {journal} {Nuclear Physics B}\ }\textbf {\bibinfo
  {volume} {431}},\ \bibinfo {pages} {484} (\bibinfo {year} {1994})},\ \Eprint
  {http://arxiv.org/abs/hep-th/9408099} {hep-th/9408099} \BibitemShut {NoStop}%
\bibitem [{\citenamefont {{Intriligator}}\ and\ \citenamefont
  {{Seiberg}}(1996)}]{1996IntriligatorSeiberg}%
  \BibitemOpen
  \bibfield  {author} {\bibinfo {author} {\bibfnamefont {K.}~\bibnamefont
  {{Intriligator}}}\ and\ \bibinfo {author} {\bibfnamefont {N.}~\bibnamefont
  {{Seiberg}}},\ }\href {\doibase 10.1016/0920-5632(95)00626-5} {\bibfield
  {journal} {\bibinfo  {journal} {Nuclear Physics B Proceedings Supplements}\
  }\textbf {\bibinfo {volume} {45}},\ \bibinfo {pages} {1} (\bibinfo {year}
  {1996})},\ \Eprint {http://arxiv.org/abs/hep-th/9509066} {hep-th/9509066}
  \BibitemShut {NoStop}%
\bibitem [{\citenamefont {{Aharony}}\ \emph {et~al.}(2000)\citenamefont
  {{Aharony}}, \citenamefont {{Gubser}}, \citenamefont {{Maldacena}},
  \citenamefont {{Ooguri}},\ and\ \citenamefont {{Oz}}}]{MAGOO}%
  \BibitemOpen
  \bibfield  {author} {\bibinfo {author} {\bibfnamefont {O.}~\bibnamefont
  {{Aharony}}}, \bibinfo {author} {\bibfnamefont {S.~S.}\ \bibnamefont
  {{Gubser}}}, \bibinfo {author} {\bibfnamefont {J.}~\bibnamefont
  {{Maldacena}}}, \bibinfo {author} {\bibfnamefont {H.}~\bibnamefont
  {{Ooguri}}}, \ and\ \bibinfo {author} {\bibfnamefont {Y.}~\bibnamefont
  {{Oz}}},\ }\href {\doibase 10.1016/S0370-1573(99)00083-6} {\bibfield
  {journal} {\bibinfo  {journal} {Physics Reports}\ }\textbf {\bibinfo {volume}
  {323}},\ \bibinfo {pages} {183} (\bibinfo {year} {2000})},\ \Eprint
  {http://arxiv.org/abs/hep-th/9905111} {hep-th/9905111} \BibitemShut {NoStop}%
\bibitem [{\citenamefont {Dasgupta}\ and\ \citenamefont
  {Halperin}(1981)}]{Dasgupta1981}%
  \BibitemOpen
  \bibfield  {author} {\bibinfo {author} {\bibfnamefont {C.}~\bibnamefont
  {Dasgupta}}\ and\ \bibinfo {author} {\bibfnamefont {B.~I.}\ \bibnamefont
  {Halperin}},\ }\href {\doibase 10.1103/PhysRevLett.47.1556} {\bibfield
  {journal} {\bibinfo  {journal} {Phys. Rev. Lett.}\ }\textbf {\bibinfo
  {volume} {47}},\ \bibinfo {pages} {1556} (\bibinfo {year}
  {1981})}\BibitemShut {NoStop}%
\bibitem [{\citenamefont {Peskin}(1978)}]{Peskin1978}%
  \BibitemOpen
  \bibfield  {author} {\bibinfo {author} {\bibfnamefont {M.~E.}\ \bibnamefont
  {Peskin}},\ }\href {\doibase 10.1016/0003-4916(78)90252-X} {\bibfield
  {journal} {\bibinfo  {journal} {Annals of Physics}\ }\textbf {\bibinfo
  {volume} {113}},\ \bibinfo {pages} {122} (\bibinfo {year}
  {1978})}\BibitemShut {NoStop}%
\bibitem [{\citenamefont {Fisher}\ and\ \citenamefont
  {Lee}(1989)}]{Fisher1989a}%
  \BibitemOpen
  \bibfield  {author} {\bibinfo {author} {\bibfnamefont {M.~P.~A.}\
  \bibnamefont {Fisher}}\ and\ \bibinfo {author} {\bibfnamefont {D.~H.}\
  \bibnamefont {Lee}},\ }\href {\doibase 10.1103/PhysRevB.39.2756} {\bibfield
  {journal} {\bibinfo  {journal} {Phys. Rev. B.}\ }\textbf {\bibinfo {volume}
  {39}},\ \bibinfo {pages} {2756} (\bibinfo {year} {1989})}\BibitemShut
  {NoStop}%
\bibitem [{\citenamefont {Naculich}\ and\ \citenamefont
  {Schnitzer}(1990)}]{Naculich:1990hg}%
  \BibitemOpen
  \bibfield  {author} {\bibinfo {author} {\bibfnamefont {S.~G.}\ \bibnamefont
  {Naculich}}\ and\ \bibinfo {author} {\bibfnamefont {H.~J.}\ \bibnamefont
  {Schnitzer}},\ }\href {\doibase 10.1016/0550-3213(90)90380-V} {\bibfield
  {journal} {\bibinfo  {journal} {Nucl. Phys.}\ }\textbf {\bibinfo {volume}
  {B347}},\ \bibinfo {pages} {687} (\bibinfo {year} {1990})}\BibitemShut
  {NoStop}%
\bibitem [{\citenamefont {Nakanishi}\ and\ \citenamefont
  {Tsuchiya}(1992)}]{Nakanishi:1990hj}%
  \BibitemOpen
  \bibfield  {author} {\bibinfo {author} {\bibfnamefont {T.}~\bibnamefont
  {Nakanishi}}\ and\ \bibinfo {author} {\bibfnamefont {A.}~\bibnamefont
  {Tsuchiya}},\ }\href {\doibase 10.1007/BF02101097} {\bibfield  {journal}
  {\bibinfo  {journal} {Commun. Math. Phys.}\ }\textbf {\bibinfo {volume}
  {144}},\ \bibinfo {pages} {351} (\bibinfo {year} {1992})}\BibitemShut
  {NoStop}%
\bibitem [{\citenamefont {Aharony}(2016)}]{Aharony2016}%
  \BibitemOpen
  \bibfield  {author} {\bibinfo {author} {\bibfnamefont {O.}~\bibnamefont
  {Aharony}},\ }\href {\doibase 10.1007/JHEP02(2016)093} {\bibfield  {journal}
  {\bibinfo  {journal} {Journal of High Energy Physics}\ }\textbf {\bibinfo
  {volume} {2016}},\ \bibinfo {pages} {93} (\bibinfo {year} {2016})},\ \Eprint
  {http://arxiv.org/abs/1512.00161} {arXiv:1512.00161} \BibitemShut {NoStop}%
\bibitem [{\citenamefont {Giombi}\ \emph {et~al.}(2012)\citenamefont {Giombi},
  \citenamefont {Minwalla}, \citenamefont {Prakash}, \citenamefont {Trivedi},
  \citenamefont {Wadia},\ and\ \citenamefont {Yin}}]{Giombi2011}%
  \BibitemOpen
  \bibfield  {author} {\bibinfo {author} {\bibfnamefont {S.}~\bibnamefont
  {Giombi}}, \bibinfo {author} {\bibfnamefont {S.}~\bibnamefont {Minwalla}},
  \bibinfo {author} {\bibfnamefont {S.}~\bibnamefont {Prakash}}, \bibinfo
  {author} {\bibfnamefont {S.~P.}\ \bibnamefont {Trivedi}}, \bibinfo {author}
  {\bibfnamefont {S.~R.}\ \bibnamefont {Wadia}}, \ and\ \bibinfo {author}
  {\bibfnamefont {X.}~\bibnamefont {Yin}},\ }\href {\doibase
  10.1140/epjc/s10052-012-2112-0} {\bibfield  {journal} {\bibinfo  {journal}
  {The European Physical Journal C}\ }\textbf {\bibinfo {volume} {72}},\
  \bibinfo {pages} {2112} (\bibinfo {year} {2012})},\ \Eprint
  {http://arxiv.org/abs/1110.4386} {arXiv:1110.4386} \BibitemShut {NoStop}%
\bibitem [{\citenamefont {{Aharony}}\ \emph {et~al.}(2012)\citenamefont
  {{Aharony}}, \citenamefont {{Gur-Ari}},\ and\ \citenamefont
  {{Yacoby}}}]{2012JHEP...03..037A}%
  \BibitemOpen
  \bibfield  {author} {\bibinfo {author} {\bibfnamefont {O.}~\bibnamefont
  {{Aharony}}}, \bibinfo {author} {\bibfnamefont {G.}~\bibnamefont
  {{Gur-Ari}}}, \ and\ \bibinfo {author} {\bibfnamefont {R.}~\bibnamefont
  {{Yacoby}}},\ }\href {\doibase 10.1007/JHEP03(2012)037} {\bibfield  {journal}
  {\bibinfo  {journal} {Journal of High Energy Physics}\ }\textbf {\bibinfo
  {volume} {3}},\ \bibinfo {eid} {37} (\bibinfo {year} {2012})},\ \Eprint
  {http://arxiv.org/abs/1110.4382} {arXiv:1110.4382 [hep-th]} \BibitemShut
  {NoStop}%
\bibitem [{\citenamefont {Aharony}\ \emph {et~al.}(2012)\citenamefont
  {Aharony}, \citenamefont {Gur-Ari},\ and\ \citenamefont
  {Yacoby}}]{Aharony2012a}%
  \BibitemOpen
  \bibfield  {author} {\bibinfo {author} {\bibfnamefont {O.}~\bibnamefont
  {Aharony}}, \bibinfo {author} {\bibfnamefont {G.}~\bibnamefont {Gur-Ari}}, \
  and\ \bibinfo {author} {\bibfnamefont {R.}~\bibnamefont {Yacoby}},\ }\href
  {\doibase 10.1007/JHEP12(2012)028} {\bibfield  {journal} {\bibinfo  {journal}
  {Journal of High Energy Physics}\ }\textbf {\bibinfo {volume} {2012}},\
  \bibinfo {pages} {28} (\bibinfo {year} {2012})},\ \Eprint
  {http://arxiv.org/abs/1207.4593} {arXiv:1207.4593} \BibitemShut {NoStop}%
\bibitem [{\citenamefont {Hsin}\ and\ \citenamefont
  {Seiberg}(2016)}]{Hsin:2016blu}%
  \BibitemOpen
  \bibfield  {author} {\bibinfo {author} {\bibfnamefont {P.-S.}\ \bibnamefont
  {Hsin}}\ and\ \bibinfo {author} {\bibfnamefont {N.}~\bibnamefont {Seiberg}},\
  }\href {\doibase 10.1007/JHEP09(2016)095} {\bibfield  {journal} {\bibinfo
  {journal} {JHEP}\ }\textbf {\bibinfo {volume} {09}},\ \bibinfo {pages} {095}
  (\bibinfo {year} {2016})},\ \Eprint {http://arxiv.org/abs/1607.07457}
  {arXiv:1607.07457 [hep-th]} \BibitemShut {NoStop}%
\bibitem [{\citenamefont {Seiberg}\ \emph {et~al.}(2016)\citenamefont
  {Seiberg}, \citenamefont {Senthil}, \citenamefont {Wang},\ and\ \citenamefont
  {Witten}}]{Seiberg2016a}%
  \BibitemOpen
  \bibfield  {author} {\bibinfo {author} {\bibfnamefont {N.}~\bibnamefont
  {Seiberg}}, \bibinfo {author} {\bibfnamefont {T.}~\bibnamefont {Senthil}},
  \bibinfo {author} {\bibfnamefont {C.}~\bibnamefont {Wang}}, \ and\ \bibinfo
  {author} {\bibfnamefont {E.}~\bibnamefont {Witten}},\ }\href {\doibase
  https://doi.org/10.1016/j.aop.2016.08.007} {\bibfield  {journal} {\bibinfo
  {journal} {Annals of Physics}\ }\textbf {\bibinfo {volume} {374}},\ \bibinfo
  {pages} {395 } (\bibinfo {year} {2016})},\ \Eprint
  {http://arxiv.org/abs/1606.01989} {arXiv:1606.01989} \BibitemShut {NoStop}%
\bibitem [{\citenamefont {Karch}\ and\ \citenamefont {Tong}(2016)}]{Karch2016}%
  \BibitemOpen
  \bibfield  {author} {\bibinfo {author} {\bibfnamefont {A.}~\bibnamefont
  {Karch}}\ and\ \bibinfo {author} {\bibfnamefont {D.}~\bibnamefont {Tong}},\
  }\href {\doibase 10.1103/PhysRevX.6.031043} {\bibfield  {journal} {\bibinfo
  {journal} {Phys. Rev. X}\ }\textbf {\bibinfo {volume} {6}},\ \bibinfo {pages}
  {031043} (\bibinfo {year} {2016})},\ \Eprint
  {http://arxiv.org/abs/1606.01893} {arXiv:1606.01893} \BibitemShut {NoStop}%
\bibitem [{\citenamefont {Wang}\ and\ \citenamefont
  {Senthil}(2016)}]{Wang2016}%
  \BibitemOpen
  \bibfield  {author} {\bibinfo {author} {\bibfnamefont {C.}~\bibnamefont
  {Wang}}\ and\ \bibinfo {author} {\bibfnamefont {T.}~\bibnamefont {Senthil}},\
  }\href {\doibase 10.1103/PhysRevB.93.085110} {\bibfield  {journal} {\bibinfo
  {journal} {Phys. Rev. B}\ }\textbf {\bibinfo {volume} {93}},\ \bibinfo
  {pages} {085110} (\bibinfo {year} {2016})},\ \Eprint
  {http://arxiv.org/abs/1507.08290} {arXiv:1507.08290} \BibitemShut {NoStop}%
\bibitem [{\citenamefont {Mross}\ \emph
  {et~al.}(2016{\natexlab{a}})\citenamefont {Mross}, \citenamefont {Alicea},\
  and\ \citenamefont {Motrunich}}]{MrossAliceaMotrunichexplicitderivation2016}%
  \BibitemOpen
  \bibfield  {author} {\bibinfo {author} {\bibfnamefont {D.~F.}\ \bibnamefont
  {Mross}}, \bibinfo {author} {\bibfnamefont {J.}~\bibnamefont {Alicea}}, \
  and\ \bibinfo {author} {\bibfnamefont {O.~I.}\ \bibnamefont {Motrunich}},\
  }\href {\doibase 10.1103/PhysRevLett.117.016802} {\bibfield  {journal}
  {\bibinfo  {journal} {Phys. Rev. Lett.}\ }\textbf {\bibinfo {volume} {117}},\
  \bibinfo {pages} {016802} (\bibinfo {year} {2016}{\natexlab{a}})}\BibitemShut
  {NoStop}%
\bibitem [{\citenamefont {Wang}\ and\ \citenamefont
  {Senthil}(2015)}]{PhysRevX.5.041031}%
  \BibitemOpen
  \bibfield  {author} {\bibinfo {author} {\bibfnamefont {C.}~\bibnamefont
  {Wang}}\ and\ \bibinfo {author} {\bibfnamefont {T.}~\bibnamefont {Senthil}},\
  }\href {\doibase 10.1103/PhysRevX.5.041031} {\bibfield  {journal} {\bibinfo
  {journal} {Phys. Rev. X}\ }\textbf {\bibinfo {volume} {5}},\ \bibinfo {pages}
  {041031} (\bibinfo {year} {2015})}\BibitemShut {NoStop}%
\bibitem [{\citenamefont {Metlitski}\ and\ \citenamefont
  {Vishwanath}(2016)}]{PhysRevB.93.245151}%
  \BibitemOpen
  \bibfield  {author} {\bibinfo {author} {\bibfnamefont {M.~A.}\ \bibnamefont
  {Metlitski}}\ and\ \bibinfo {author} {\bibfnamefont {A.}~\bibnamefont
  {Vishwanath}},\ }\href {\doibase 10.1103/PhysRevB.93.245151} {\bibfield
  {journal} {\bibinfo  {journal} {Phys. Rev. B}\ }\textbf {\bibinfo {volume}
  {93}},\ \bibinfo {pages} {245151} (\bibinfo {year} {2016})}\BibitemShut
  {NoStop}%
\bibitem [{\citenamefont {Xu}\ and\ \citenamefont
  {You}(2015)}]{XuYou2015selfdual}%
  \BibitemOpen
  \bibfield  {author} {\bibinfo {author} {\bibfnamefont {C.}~\bibnamefont
  {Xu}}\ and\ \bibinfo {author} {\bibfnamefont {Y.-Z.}\ \bibnamefont {You}},\
  }\href {\doibase 10.1103/PhysRevB.92.220416} {\bibfield  {journal} {\bibinfo
  {journal} {Phys. Rev. B}\ }\textbf {\bibinfo {volume} {92}},\ \bibinfo
  {pages} {220416} (\bibinfo {year} {2015})}\BibitemShut {NoStop}%
\bibitem [{\citenamefont {Son}(2015)}]{PhysRevX.5.031027}%
  \BibitemOpen
  \bibfield  {author} {\bibinfo {author} {\bibfnamefont {D.~T.}\ \bibnamefont
  {Son}},\ }\href {\doibase 10.1103/PhysRevX.5.031027} {\bibfield  {journal}
  {\bibinfo  {journal} {Phys. Rev. X}\ }\textbf {\bibinfo {volume} {5}},\
  \bibinfo {pages} {031027} (\bibinfo {year} {2015})}\BibitemShut {NoStop}%
\bibitem [{\citenamefont {Mulligan}\ \emph {et~al.}(2016)\citenamefont
  {Mulligan}, \citenamefont {Raghu},\ and\ \citenamefont
  {Fisher}}]{PhysRevB.94.075101}%
  \BibitemOpen
  \bibfield  {author} {\bibinfo {author} {\bibfnamefont {M.}~\bibnamefont
  {Mulligan}}, \bibinfo {author} {\bibfnamefont {S.}~\bibnamefont {Raghu}}, \
  and\ \bibinfo {author} {\bibfnamefont {M.~P.~A.}\ \bibnamefont {Fisher}},\
  }\href {\doibase 10.1103/PhysRevB.94.075101} {\bibfield  {journal} {\bibinfo
  {journal} {Phys. Rev. B}\ }\textbf {\bibinfo {volume} {94}},\ \bibinfo
  {pages} {075101} (\bibinfo {year} {2016})}\BibitemShut {NoStop}%
\bibitem [{\citenamefont {Sodemann}\ \emph {et~al.}(2017)\citenamefont
  {Sodemann}, \citenamefont {Kimchi}, \citenamefont {Wang},\ and\ \citenamefont
  {Senthil}}]{PhysRevB.95.085135}%
  \BibitemOpen
  \bibfield  {author} {\bibinfo {author} {\bibfnamefont {I.}~\bibnamefont
  {Sodemann}}, \bibinfo {author} {\bibfnamefont {I.}~\bibnamefont {Kimchi}},
  \bibinfo {author} {\bibfnamefont {C.}~\bibnamefont {Wang}}, \ and\ \bibinfo
  {author} {\bibfnamefont {T.}~\bibnamefont {Senthil}},\ }\href {\doibase
  10.1103/PhysRevB.95.085135} {\bibfield  {journal} {\bibinfo  {journal} {Phys.
  Rev. B}\ }\textbf {\bibinfo {volume} {95}},\ \bibinfo {pages} {085135}
  (\bibinfo {year} {2017})}\BibitemShut {NoStop}%
\bibitem [{\citenamefont {Kachru}\ \emph {et~al.}(2015)\citenamefont {Kachru},
  \citenamefont {Mulligan}, \citenamefont {Torroba},\ and\ \citenamefont
  {Wang}}]{PhysRevB.92.235105}%
  \BibitemOpen
  \bibfield  {author} {\bibinfo {author} {\bibfnamefont {S.}~\bibnamefont
  {Kachru}}, \bibinfo {author} {\bibfnamefont {M.}~\bibnamefont {Mulligan}},
  \bibinfo {author} {\bibfnamefont {G.}~\bibnamefont {Torroba}}, \ and\
  \bibinfo {author} {\bibfnamefont {H.}~\bibnamefont {Wang}},\ }\href {\doibase
  10.1103/PhysRevB.92.235105} {\bibfield  {journal} {\bibinfo  {journal} {Phys.
  Rev. B}\ }\textbf {\bibinfo {volume} {92}},\ \bibinfo {pages} {235105}
  (\bibinfo {year} {2015})}\BibitemShut {NoStop}%
\bibitem [{\citenamefont {{Geraedts}}\ \emph {et~al.}(2016)\citenamefont
  {{Geraedts}}, \citenamefont {{Zaletel}}, \citenamefont {{Mong}},
  \citenamefont {{Metlitski}}, \citenamefont {{Vishwanath}},\ and\
  \citenamefont {{Motrunich}}}]{Geraedtsetal2015}%
  \BibitemOpen
  \bibfield  {author} {\bibinfo {author} {\bibfnamefont {S.~D.}\ \bibnamefont
  {{Geraedts}}}, \bibinfo {author} {\bibfnamefont {M.~P.}\ \bibnamefont
  {{Zaletel}}}, \bibinfo {author} {\bibfnamefont {R.~S.~K.}\ \bibnamefont
  {{Mong}}}, \bibinfo {author} {\bibfnamefont {M.~A.}\ \bibnamefont
  {{Metlitski}}}, \bibinfo {author} {\bibfnamefont {A.}~\bibnamefont
  {{Vishwanath}}}, \ and\ \bibinfo {author} {\bibfnamefont {O.~I.}\
  \bibnamefont {{Motrunich}}},\ }\href {\doibase 10.1126/science.aad4302}
  {\bibfield  {journal} {\bibinfo  {journal} {Science}\ }\textbf {\bibinfo
  {volume} {352}},\ \bibinfo {pages} {197} (\bibinfo {year} {2016})},\ \Eprint
  {http://arxiv.org/abs/1508.04140} {arXiv:1508.04140 [cond-mat.str-el]}
  \BibitemShut {NoStop}%
\bibitem [{\citenamefont {Mross}\ \emph
  {et~al.}(2016{\natexlab{b}})\citenamefont {Mross}, \citenamefont {Alicea},\
  and\ \citenamefont {Motrunich}}]{PhysRevLett.117.136802}%
  \BibitemOpen
  \bibfield  {author} {\bibinfo {author} {\bibfnamefont {D.~F.}\ \bibnamefont
  {Mross}}, \bibinfo {author} {\bibfnamefont {J.}~\bibnamefont {Alicea}}, \
  and\ \bibinfo {author} {\bibfnamefont {O.~I.}\ \bibnamefont {Motrunich}},\
  }\href {\doibase 10.1103/PhysRevLett.117.136802} {\bibfield  {journal}
  {\bibinfo  {journal} {Phys. Rev. Lett.}\ }\textbf {\bibinfo {volume} {117}},\
  \bibinfo {pages} {136802} (\bibinfo {year} {2016}{\natexlab{b}})}\BibitemShut
  {NoStop}%
\bibitem [{\citenamefont {Fisher}\ \emph {et~al.}(1989)\citenamefont {Fisher},
  \citenamefont {Weichman}, \citenamefont {Grinstein},\ and\ \citenamefont
  {Fisher}}]{Fisher1989}%
  \BibitemOpen
  \bibfield  {author} {\bibinfo {author} {\bibfnamefont {M.~P.~A.}\
  \bibnamefont {Fisher}}, \bibinfo {author} {\bibfnamefont {P.~B.}\
  \bibnamefont {Weichman}}, \bibinfo {author} {\bibfnamefont {G.}~\bibnamefont
  {Grinstein}}, \ and\ \bibinfo {author} {\bibfnamefont {D.~S.}\ \bibnamefont
  {Fisher}},\ }\href {\doibase 10.1103/PhysRevB.40.546} {\bibfield  {journal}
  {\bibinfo  {journal} {Phys. Rev. B.}\ }\textbf {\bibinfo {volume} {40}},\
  \bibinfo {pages} {546} (\bibinfo {year} {1989})}\BibitemShut {NoStop}%
\bibitem [{Note1()}]{Note1}%
  \BibitemOpen
  \bibinfo {note} {One can legalize the theory by introducing a new dynamical
  gauge field $b$ as a constraint, giving us the Lagrangian \begin {multline*}
  \protect \mathcal {L}= |(\partial _\mu -ie^*A_\mu - ia_\mu )\phi |^2 + g|\phi
  |^4 \\ + \protect \frac {i}{4\pi }\epsilon ^{\mu \nu \lambda } \left (-k^B
  b_\mu \partial _\nu b_\lambda + 2 a_\mu \partial _\nu b_\lambda \right ) \end
  {multline*}}\BibitemShut {NoStop}%
\bibitem [{\citenamefont {Tong}(2016)}]{Tong2016}%
  \BibitemOpen
  \bibfield  {author} {\bibinfo {author} {\bibfnamefont {D.}~\bibnamefont
  {Tong}},\ }\href {http://arxiv.org/abs/1606.06687} {\enquote {\bibinfo
  {title} {{Lectures on the Quantum Hall Effect}},}\ } (\bibinfo {year}
  {2016}),\ \Eprint {http://arxiv.org/abs/1606.06687} {arXiv:1606.06687}
  \BibitemShut {NoStop}%
\bibitem [{\citenamefont {Lee}(1991)}]{Lee1991}%
  \BibitemOpen
  \bibfield  {author} {\bibinfo {author} {\bibfnamefont {D.-H.}\ \bibnamefont
  {Lee}},\ }\href {\doibase 10.1142/S0217979291001607} {\bibfield  {journal}
  {\bibinfo  {journal} {International Journal of Modern Physics B}\ }\textbf
  {\bibinfo {volume} {05}},\ \bibinfo {pages} {1695} (\bibinfo {year}
  {1991})}\BibitemShut {NoStop}%
\bibitem [{\citenamefont {Giombi}\ \emph {et~al.}(2017)\citenamefont {Giombi},
  \citenamefont {Gurucharan}, \citenamefont {Kirilin}, \citenamefont
  {Prakash},\ and\ \citenamefont {Skvortsov}}]{Giombi2016}%
  \BibitemOpen
  \bibfield  {author} {\bibinfo {author} {\bibfnamefont {S.}~\bibnamefont
  {Giombi}}, \bibinfo {author} {\bibfnamefont {V.}~\bibnamefont {Gurucharan}},
  \bibinfo {author} {\bibfnamefont {V.}~\bibnamefont {Kirilin}}, \bibinfo
  {author} {\bibfnamefont {S.}~\bibnamefont {Prakash}}, \ and\ \bibinfo
  {author} {\bibfnamefont {E.}~\bibnamefont {Skvortsov}},\ }\href {\doibase
  10.1007/JHEP01(2017)058} {\bibfield  {journal} {\bibinfo  {journal} {Journal
  of High Energy Physics}\ }\textbf {\bibinfo {volume} {2017}},\ \bibinfo
  {pages} {58} (\bibinfo {year} {2017})},\ \Eprint
  {http://arxiv.org/abs/1610.08472} {arXiv:1610.08472} \BibitemShut {NoStop}%
\bibitem [{\citenamefont {Aharony}\ \emph {et~al.}(2013)\citenamefont
  {Aharony}, \citenamefont {Giombi}, \citenamefont {Gur-Ari}, \citenamefont
  {Maldacena},\ and\ \citenamefont {Yacoby}}]{Aharony2012}%
  \BibitemOpen
  \bibfield  {author} {\bibinfo {author} {\bibfnamefont {O.}~\bibnamefont
  {Aharony}}, \bibinfo {author} {\bibfnamefont {S.}~\bibnamefont {Giombi}},
  \bibinfo {author} {\bibfnamefont {G.}~\bibnamefont {Gur-Ari}}, \bibinfo
  {author} {\bibfnamefont {J.}~\bibnamefont {Maldacena}}, \ and\ \bibinfo
  {author} {\bibfnamefont {R.}~\bibnamefont {Yacoby}},\ }\href {\doibase
  10.1007/JHEP03(2013)121} {\bibfield  {journal} {\bibinfo  {journal} {Journal
  of High Energy Physics}\ }\textbf {\bibinfo {volume} {2013}},\ \bibinfo
  {pages} {121} (\bibinfo {year} {2013})},\ \Eprint
  {http://arxiv.org/abs/1211.4843} {arXiv:1211.4843} \BibitemShut {NoStop}%
\bibitem [{Note2()}]{Note2}%
  \BibitemOpen
  \bibinfo {note} {By dimensional regularization, we mean that one contracts
  tensor indices in 3 dimensions, while analytically continuing integrals to
  $3-\epsilon $ dimensions. This is sometimes called dimensional reduction in
  the literature.\cite {Chen1992} An alternative scheme where one regularizes
  with a small Yang-Mills term is equivalent to dimensional regularization, up
  to a constant shift of the $SU(N)$ level, so we will work exclusively in
  dimensional regularization for simplicity.}\BibitemShut {Stop}%
\bibitem [{\citenamefont {Gurucharan}\ and\ \citenamefont
  {Prakash}(2014)}]{Gurucharan2014}%
  \BibitemOpen
  \bibfield  {author} {\bibinfo {author} {\bibfnamefont {V.}~\bibnamefont
  {Gurucharan}}\ and\ \bibinfo {author} {\bibfnamefont {S.}~\bibnamefont
  {Prakash}},\ }\href {\doibase 10.1007/JHEP09(2014)009} {\bibfield  {journal}
  {\bibinfo  {journal} {Journal of High Energy Physics}\ }\textbf {\bibinfo
  {volume} {2014}},\ \bibinfo {pages} {9} (\bibinfo {year} {2014})},\ \Eprint
  {http://arxiv.org/abs/1404.7849} {arXiv:1404.7849} \BibitemShut {NoStop}%
\bibitem [{\citenamefont {Chester}\ and\ \citenamefont
  {Pufu}(2016)}]{Chester2016}%
  \BibitemOpen
  \bibfield  {author} {\bibinfo {author} {\bibfnamefont {S.~M.}\ \bibnamefont
  {Chester}}\ and\ \bibinfo {author} {\bibfnamefont {S.~S.}\ \bibnamefont
  {Pufu}},\ }\href {\doibase 10.1007/JHEP08(2016)069} {\bibfield  {journal}
  {\bibinfo  {journal} {Journal of High Energy Physics}\ }\textbf {\bibinfo
  {volume} {2016}},\ \bibinfo {pages} {69} (\bibinfo {year} {2016})},\ \Eprint
  {http://arxiv.org/abs/1603.05582} {arXiv:1603.05582} \BibitemShut {NoStop}%
\bibitem [{Note3()}]{Note3}%
  \BibitemOpen
  \bibinfo {note} {This conclusion might be further supported by the fact that
  the bosonization dual of the model studied by Chen, Fisher, and Wu involves a
  boson coupled to a Chern-Simons gauge field with non-abelian gauge group,
  rather than an abelian gauge field. In particular, \begin {gather*}
  U(1)_{k^B-1/2, k^B-1/2} \protect \text { with one fermion} \\ \protect
  \updownarrow \\ U(k^B-1)_{-k^B, -k^B} \protect \text { with one boson} \end
  {gather*}.}\BibitemShut {Stop}%
\bibitem [{\citenamefont {{Hui}}\ \emph {et~al.}(2017)\citenamefont {{Hui}},
  \citenamefont {{Kim}},\ and\ \citenamefont
  {{Mulligan}}}]{2017arXiv171204942H}%
  \BibitemOpen
  \bibfield  {author} {\bibinfo {author} {\bibfnamefont {A.}~\bibnamefont
  {{Hui}}}, \bibinfo {author} {\bibfnamefont {E.-A.}\ \bibnamefont {{Kim}}}, \
  and\ \bibinfo {author} {\bibfnamefont {M.}~\bibnamefont {{Mulligan}}},\
  }\href@noop {} {\bibfield  {journal} {\bibinfo  {journal} {ArXiv e-prints}\ }
  (\bibinfo {year} {2017})},\ \Eprint {http://arxiv.org/abs/1712.04942}
  {arXiv:1712.04942 [cond-mat.str-el]} \BibitemShut {NoStop}%
\bibitem [{\citenamefont {Chen}\ \emph {et~al.}(1992)\citenamefont {Chen},
  \citenamefont {Semenoff},\ and\ \citenamefont {Wu}}]{Chen1992}%
  \BibitemOpen
  \bibfield  {author} {\bibinfo {author} {\bibfnamefont {W.}~\bibnamefont
  {Chen}}, \bibinfo {author} {\bibfnamefont {G.~W.}\ \bibnamefont {Semenoff}},
  \ and\ \bibinfo {author} {\bibfnamefont {Y.-S.}\ \bibnamefont {Wu}},\ }\href
  {\doibase 10.1103/PhysRevD.46.5521} {\bibfield  {journal} {\bibinfo
  {journal} {Physical Review D}\ }\textbf {\bibinfo {volume} {46}},\ \bibinfo
  {pages} {5521} (\bibinfo {year} {1992})},\ \Eprint
  {http://arxiv.org/abs/9209005} {arXiv:9209005 [hep-th]} \BibitemShut
  {NoStop}%
\end{thebibliography}%

\appendix*
\onecolumngrid
\section{Calculational Details}
The Lagrangian we study is
\begin{equation}
    \lagr = \sum_{i=1}^{N_f} \bar{\psi}_i\gamma^\mu(\partial_\mu - ia_\mu)\psi_i + \frac{ik^F}{4\pi}\epsilon^{\mu\nu\lambda}\text{Tr}\left( a_\mu\partial_\nu a_\lambda + \frac{2}{3}a_\mu a_\nu a_\lambda\right)
\end{equation}

Define light-cone coordinates via analytic continuation to be $x^{\pm} = (x^1 \pm ix^2)/\sqrt{2}$, and let $x_s^2 = x_1^2 + x_2^2 = 2x_+x_-$. We will work in light-cone gauge $a_- = 0$, which decouples the ghosts and removes the cubic gauge interaction term.\cite{Gurucharan2014} We will also take $\gamma^i = \sigma^i$, the Pauli matrices. We normalize our gauge group generators by $\Tr T^a T^b = \delta^{ab}/2$.

We will regularize our theory by using a momentum-cutoff $\Lambda$ in the 1-2 plane and dimensional-regularization in the $x^3$ direction, as has been done by others.\cite{Aharony2012a, Gurucharan2014}

The Feynman rules for the bare propagators and interactions are

\def\arraystretch{2}
\setlength\tabcolsep{3em}
\begin{center}
\includegraphics[width=.6\columnwidth]{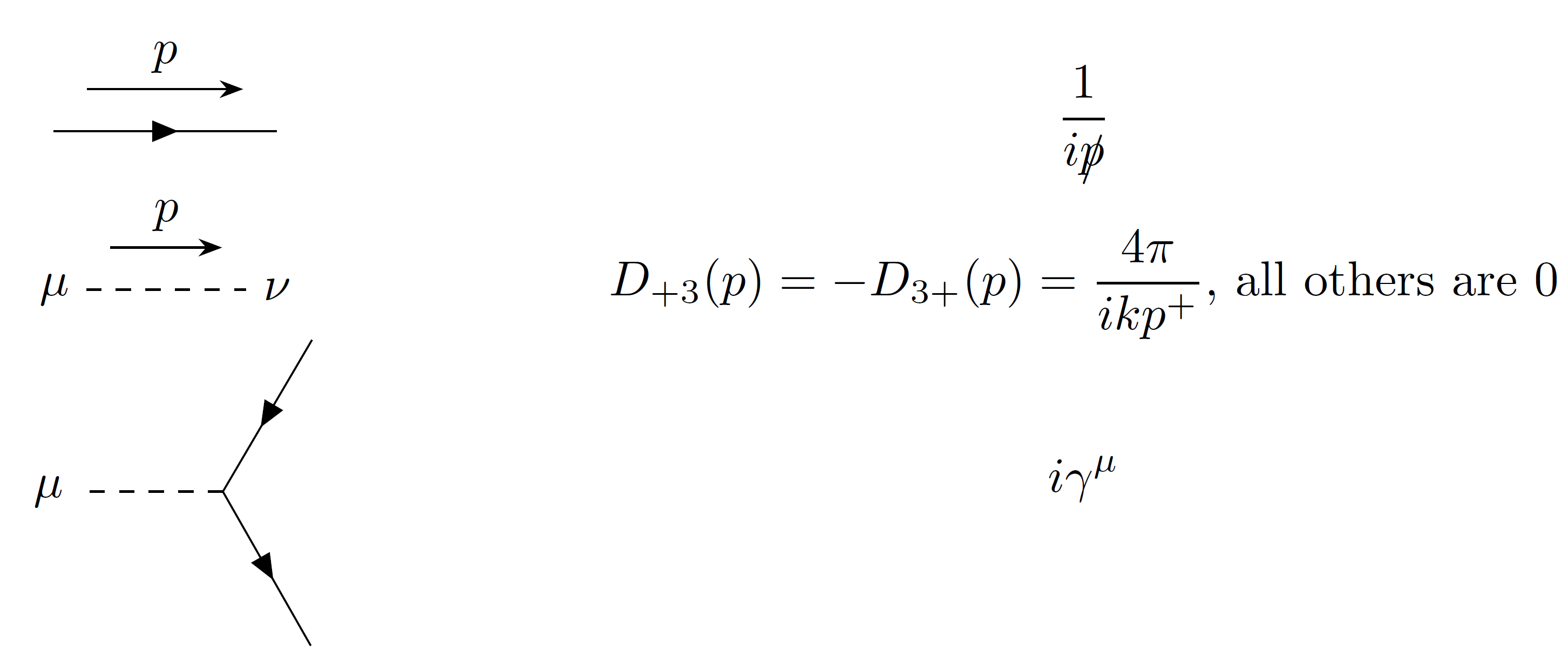}
\end{center}

Under duality, we expect $\phi^\dagger\phi \leftrightarrow \bar{\psi}\psi$. Hence, $\nu^{-1} = 3-[\bar{\psi}\psi]$.\cite{Wen1993} In what follows, we will be calculating the scaling dimension $[\bar{\psi}\psi]$.

Denote the mass operator in momentum space as $J_0(p) = (\bar{\psi}\psi) (p)$, where $p$ is the momentum inserted into the vertex. The leading order in $\alpha$ term of $\langle J_0(p) J_0(-p) \rangle \sim p$, and we know that the leading order scaling dimension $\Delta^{(0)}$ of the mass operator $J_0$ is given by $\langle J_0(p) J_0 (-p)\rangle_\text{leading} \sim p^{2\Delta^{(0)} - d}$, where $d$ is the number of spacetime dimensions. Hence, the scaling dimension of $J_0$ at leading order in (2+1)D is $2$. We will calculate the anomalous dimension $\delta^{(1)}$ of $J_0$, which amounts to extracting the logarithmic divergences of the 2-point function as\cite{Chester2016}
\begin{equation}
\langle J_0 J_0\rangle = (1-2\delta^{(1)}\ln \Lambda + \ldots) \langle J_0 J_0 \rangle_\text{leading}
\end{equation}

First, let us calculate the exact gauge propagator $G_{\mu\nu}$ to leading order in $\alpha$, which we denote by a squiggle. The only diagrams that contribute are strings of bubble diagrams, and hence satisfies the following Schwinger-Dyson equation

\begin{equation}
\includegraphics[width=.5\columnwidth]{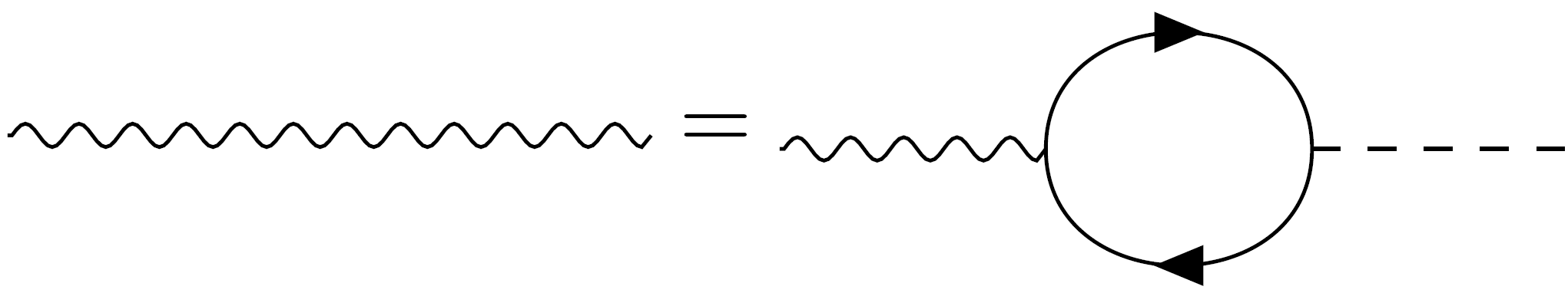}
\end{equation}

The 1PI self-energy diagram $\Sigma^{\mu\nu}$ at leading order is given by
\begin{eqnarray}
    \Sigma^{\mu\nu}(p) &=& (-1)\Tr(T^a T^b)\delta^{ab} \Tr \int\frac{d^3 q}{(2\pi)^3}\frac{-i\slashed{q}}{q^2} (i\gamma^\mu)\frac{-i(\slashed{p}+\slashed{q})}{(p+q)^2}(i\gamma^\nu) \nonumber \\ 
    &=& -\frac{N_f p}{32}\left(\delta^{\mu\nu} -\frac{p^\mu p^\nu}{p^2}\right)
\end{eqnarray}

Summing the bubbles via $\mathbf{G}(p) = (1-\mathbf{D}\mathbf{\Sigma})^{-1}\mathbf{D}(p)$, we get
\begin{equation}
    \begin{pmatrix}
        G_{33} & G_{3+}\\
        G_{+3} & G_{++}
    \end{pmatrix}(p)
    =
    \frac{1}{N_f}\frac{2\pi^2 p_+^2}{p p_s^4}\frac{64}{64+\pi^2\lambda^2}
    \begin{pmatrix}
        \lambda^2 p_-^2 & \frac{8i\lambda}{\pi}p_-p - \lambda^2p_-p_3 \\
        -\frac{8i\lambda}{\pi}p_-p - \lambda^2p_-p_3 & -p_s^2\lambda^2
    \end{pmatrix}
\end{equation}

There are four diagrams at subleading order in $\alpha$ that contribute to $\langle J_0 J_0 \rangle$. We denote a $J_0$ insertion by a crossed circle.

First, the fermion self-energy contribution. 

\begin{center}
\includegraphics[width = .4\columnwidth]{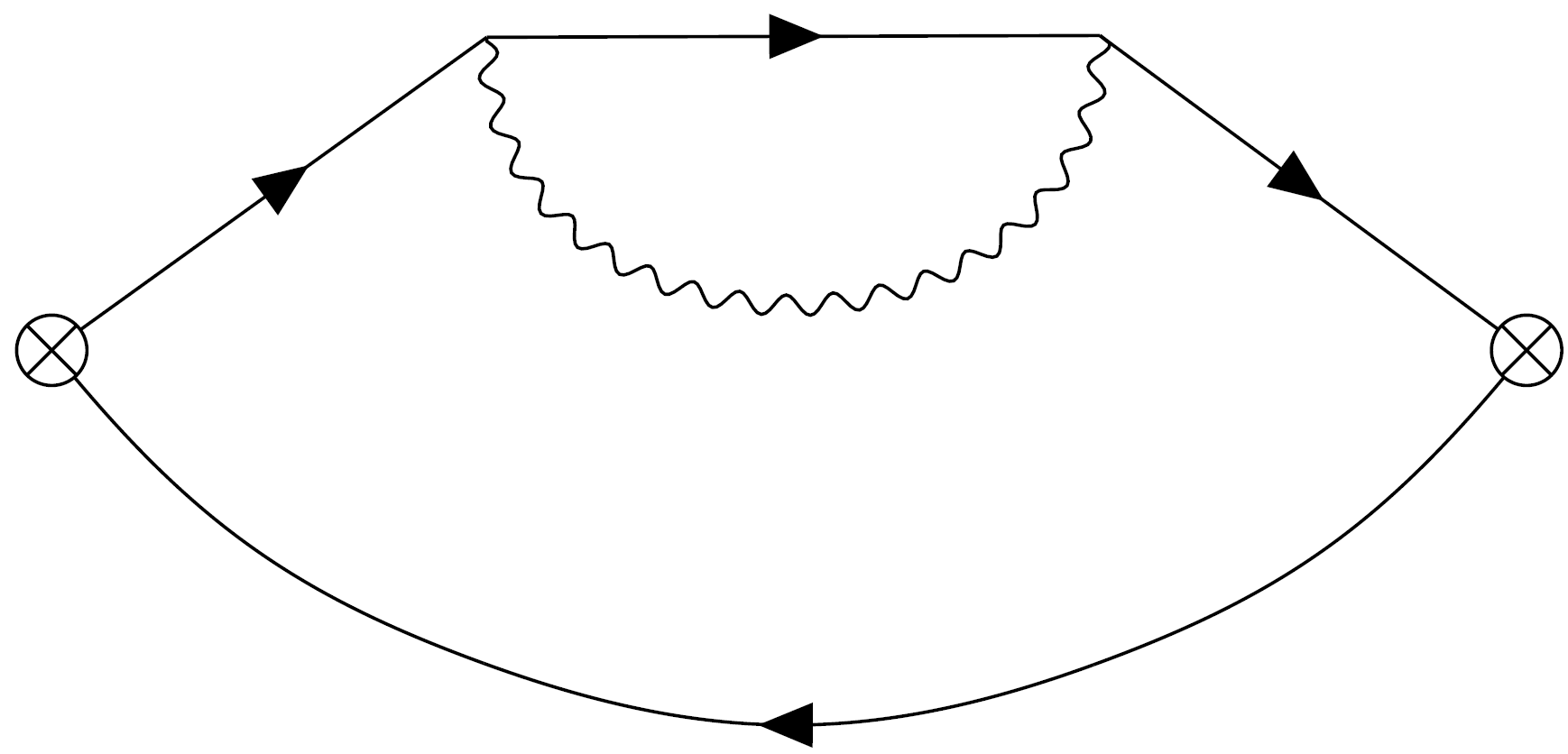}
\end{center}
We focus on the fermion self-energy subdiagram first.
\begin{equation}
    \Sigma_\psi(p) = \frac{N}{2} \int \frac{d^3 q}{(2\pi)^3}(i\gamma^\mu)\frac{-i(\slashed{p} + \slashed{q})}{(p+q)^2}(i\gamma^\nu)G_{\mu\nu}(q)
\end{equation}
Using the relations $\gamma^+ \gamma^- = 1 + \gamma^3$, $\gamma^-\gamma^+= 1-\gamma^3$, and $(\gamma^3)^2 = 1$, we get that
\begin{equation}
    \gamma^\mu\slashed{p}\gamma^\nu G_{\mu\nu} = G_{33}(p_3\gamma^3 - p_-\gamma^- -p_+\gamma^+) + (G_{+3} + G_{3+})(p_3\gamma^+ + p_- \gamma^3) + (G_{+3}-G_{3+})p_- + 2G_{++} p_-\gamma^+
\end{equation}
Substituting $\slashed{p}\rightarrow \slashed{p} + \slashed{q}$ in the above equation, we get
\begin{equation}
\Sigma_\psi(p) = i\alpha\frac{64\pi^2}{64+\pi^2\lambda^2}\int\frac{d^3q}{(2\pi)^3}\frac{1}{(p+q)^2}(K_\mu\gamma^\mu + K_I)
\end{equation}
where
\begin{eqnarray}
    K_- &=& -\frac{p_-+q_-}{4q}\lambda^2 \\
    K_+ &=& -\frac{p_+ + q_+}{4q}\lambda^2 -(p_3+q_3)\frac{q_+q_3}{qq_s^2}\lambda^2 - 2p_-\frac{q_+^2}{qq_s^2}\lambda^2 -\frac{q_+}{q}\lambda^2 \\
    K_3 &=& \frac{p_3+q_3}{4q}\lambda^2 - p_-\frac{q_3q_+}{qq_s^2}\lambda^2 - \frac{q_3}{2q}\lambda^2 \\
    K_I &=& -p_-\frac{8i}{\pi}\frac{q_+}{q_s^2}\lambda - \frac{4i}{\pi}\lambda
\end{eqnarray}

We use Feynman parameters to evaluate these integrals, and we will only keep the logarithmic divergences. The relevant formulas are
\begin{align}
     \int\frac{d^3 q}{(2\pi)^3} \frac{f(q)}{q(p+q)^2} &= \frac{1}{2}\int_0^1 dx \int\frac{d^3 q}{(2\pi)^3} (1-x)^{-1/2} \frac{f(q-xp)}{(q^2 + x(1-x)p^2)^{3/2}} \\
     \int\frac{d^3 q}{(2\pi)^3} \frac{f(q)}{q^2(p+q)^2} &= \int_0^1 dx \int\frac{d^3 q}{(2\pi)^3} \frac{f(q-xp)}{(q^2 + x(1-x)p^2} \\
     \int\frac{d^3 q}{(2\pi)^3} \frac{f(q_3, \vec{q}_s)}{q_s^2q(p+q)^2} &= \frac{3}{4}\int_0^1 dy \int_0^{1-y} dz \int \frac{d^3 q}{(2\pi)^3} y^{-1/2} \frac{f(q_3 - \frac{z}{y+z}p_3, \vec{q}_s - z\vec{p}_s)}{\left(q_s^2 + z(1-z)p_s^2 + (y+z)q_3^2 + \frac{yz}{y+z}p_3^2\right)^{5/2}}
\end{align}
The result for the fermionic self-energy is
\begin{equation}
    \Sigma_\psi(p) = i \alpha \frac{64}{64+\pi^2\lambda^2}\frac{\lambda^2}{24}(-p_\mu\gamma^\mu + 6p_3\gamma^3+12p_+\gamma^+)\ln \Lambda + \ldots
\end{equation}
Putting this into the two point function at zero external momenta, we can extract the logarithmic contribution via 
\begin{align}
    \frac{1}{2} \Tr \frac{\slashed{p}}{ip^2}\Sigma_\psi(p) &=  \frac{1}{2} \Tr \frac{\slashed{p}}{ip^2}i \alpha \frac{64}{64+\pi^2\lambda^2}\frac{\lambda^2}{24}(-p_\mu\gamma^\mu + 6p_3\gamma^3+12p_+\gamma^+)\ln \Lambda \\
    &=\alpha \frac{64\lambda^2}{64+\pi^2 \lambda^2} \frac{\lambda^2}{24} (-p^2 +6 p_3^2 + 6 p_s^2) \frac{1}{p^2} \ln \Lambda \\
    &=\alpha \frac{64\lambda^2}{64+\pi^2\lambda^2} \frac{5}{24} \ln \Lambda
\end{align}
Since this diagram contributes with a weight of $2$, it contributes $\delta_1 = -\alpha \frac{64\lambda^2}{64+\pi^2\lambda^2} \frac{5}{24}$

Next, the 1-loop vertex correction. 

\begin{center}
\includegraphics[width=.4\columnwidth]{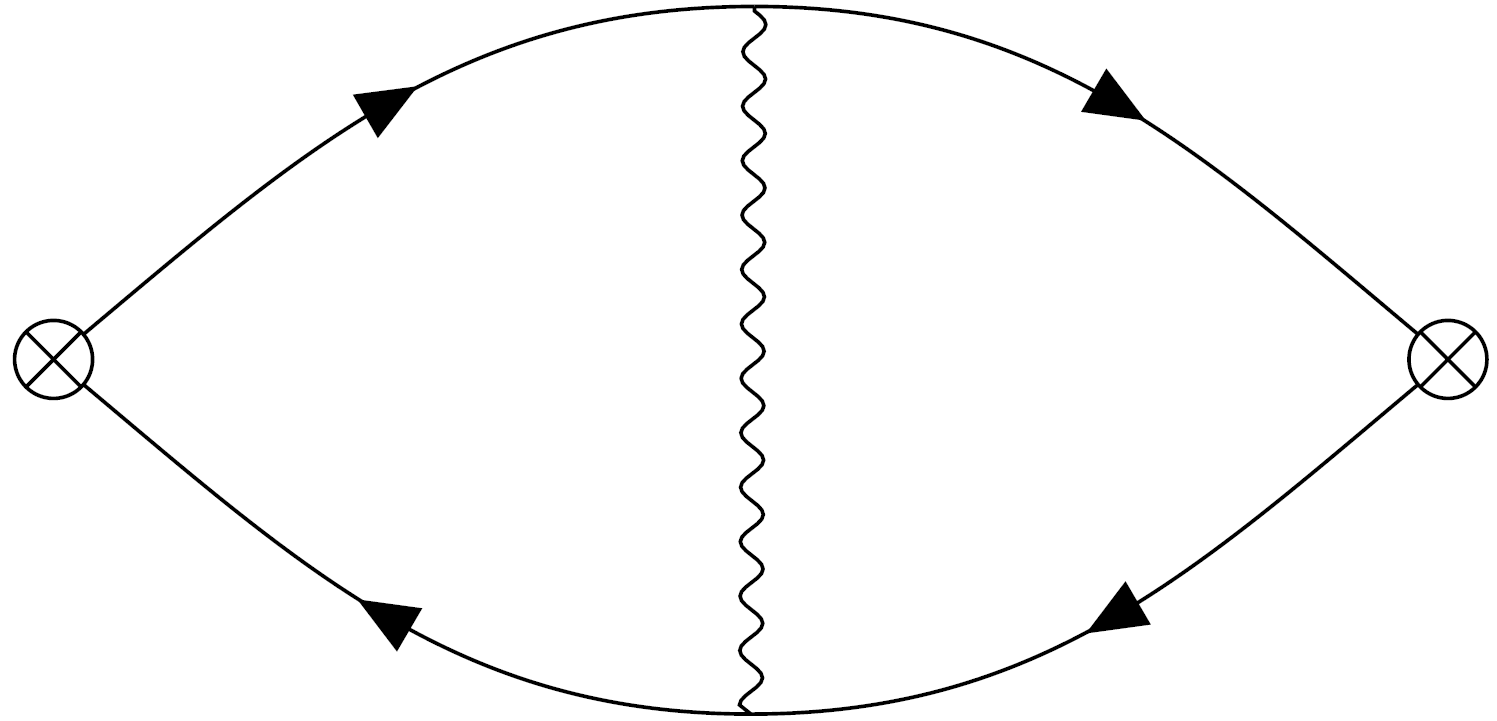}
\end{center}

Note that to extract logarithmic divergences it is easier to calculate the vertex correction with external momenta $0$ than to calculate the full two-loop integral. Also, since we will combine the two free ends to a single vertex, we only care about the identity component, which can be extracted by applying $1/2 \Tr$ over the gamma matrices. Hence, the divergence is given by

\begin{equation}
    \frac{N}{2} \frac{1}{2} \Tr \int \frac{d^3 q}{(2\pi)^3} (i\gamma^\mu)\frac{1}{-q^2}(i\gamma^\nu)G_{\mu\nu}(q) = \alpha\frac{\lambda^2}{64+\pi^2 \lambda^2} \frac{1}{8} \ln \Lambda
\end{equation}

Each vertex contributes once to the divergence, so there is an overall factor of $2$. In total, this diagram contributes $\delta_2 = - \alpha\frac{64\lambda^2}{64+\pi^2\lambda^2} \frac{1}{8}$.

Finally, the last diagrams 

\begin{center}
\includegraphics[width=.9\columnwidth]{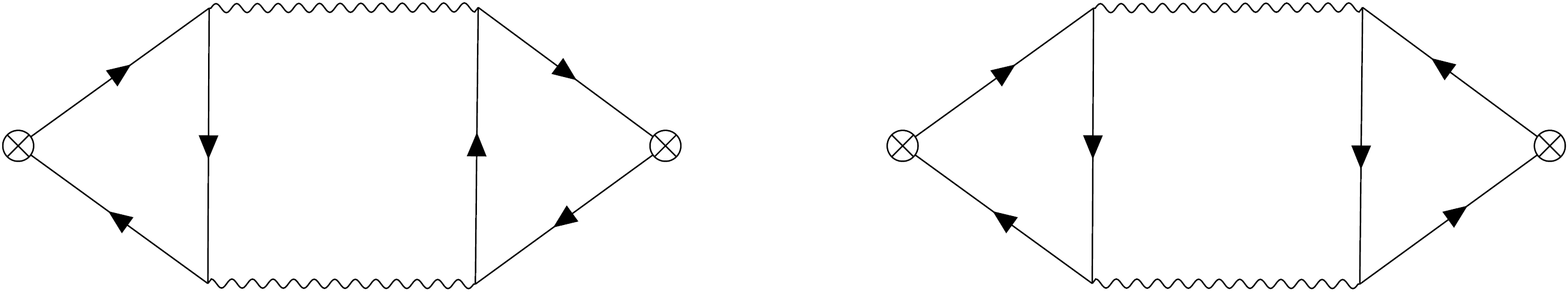}
\end{center}

These are two-loop vertex corrections, so again it's simpler to focus only on the vertex. Note that since we will combine the two free ends to a single vertex, we only care about the identity component, which can be extracted by applying $1/2 \Tr$ over the gamma matrices. We focus first on the left one.

\begin{align}
    (-1) N_f \frac{N}{4} \int \frac{d^3 p}{(2\pi)^3} \int \frac{d^3 k}{(2\pi)^3}\frac{1}{2}\Tr\left(\frac{1}{i\slashed{p}} \frac{1}{i\slashed{p}}\gamma^\sigma \frac{1}{i(\slashed{p}+\slashed{k})} \gamma^\nu\right) \Tr\left(\gamma^\mu \frac{1}{i\slashed{k}}\gamma^\eta \right) G_{\mu\nu}(k) G_{\sigma\eta}(k) 
     = \alpha \frac{1}{2} \frac{64}{64+\pi^2 \lambda^2} \frac{64-\pi^2 \lambda^2}{64+\pi^2\lambda^2} \ln \Lambda
\end{align}

This diagram contributes with a factor of $2$ because there are two vertices. The right diagram also gives the same result because of the relation $\Tr \gamma^\alpha \gamma^\beta \gamma^\delta = - \Tr \gamma^\gamma \gamma^\beta \gamma^\alpha$. Hence, the two diagrams together contribute $\delta_3 = - \alpha\frac{64}{64+\pi^2 \lambda^2} \frac{64-\pi^2\lambda^2}{64+\pi^2\lambda^2}$.

Therefore, the scaling dimension of $\bar{\psi}\psi$ is
\begin{equation}
    [\bar{\psi}\psi] = 2 - (\delta_1+\delta_2+\delta_3) = 2 - \alpha\frac{128\lambda^2}{3}\frac{128-\pi^2\lambda^2}{(64+\pi^2\lambda^2)^2}
\end{equation}
Note that our answer differs with Gurucharan and Prakash, as they did not include the last diagram which contributes an extra factor of $2$ in $\delta_3$.

\end{document}